\documentclass[3p,twocolumn]{elsarticle}

\usepackage{amsmath, amssymb, amsthm}
\usepackage{graphicx}
\usepackage{hyperref}
\usepackage{subcaption}
\usepackage{physics} 

\begin{document}
\begin{frontmatter}

\title{Mixed state concurrence for symmetric systems}
\author{S. H. Curnoe, D. Gajera, and C. Wei}
\address{Department of Physics and Physical Oceanography, Memorial University of Newfoundland, St. John's, Newfoundland and Labrador, Canada}

\date{May 2025}

\begin{abstract}
We present a method to quantify entanglement in mixed states of highly
symmetric systems. Symmetry constrains interactions between parts
and predicts the degeneracies of the states.  While symmetry alone
produces entangled eigenstates, the thermal mixed state (density) which contains
all of the eigenstate densities weighted by their Boltzmann factors is not
necessarily  as entangled as the eigenstates themselves because generally
the mixed state can be re-expressed as a sum over densities which are less
entangled.  The entanglement of the mixed state is the minimum obtained by
considering all such re-expressions, but there is no well-defined approach 
to solving this problem generally.  Our method uses symmetry to explicitly construct
unentangled densities, which are then optimally
included in the thermal mixed state, resulting in a quantitative measure of
entanglement that accounts for the reduction of entanglement
arising from degenerate states.
We present results for several small spin systems.
\end{abstract}

\end{frontmatter}

\section{Introduction}

Entanglement is a fundamental property of  quantum systems, but its quantitative evaluation poses several challenges.  The basic idea is that a measurement of one part of an entangled system will affect subsequent measurements of other parts, thus a  
pure state that is unentangled is one that is {\em separable}, that is, it can be factored, 
$|\Psi\rangle = |\psi\rangle \otimes |\phi\rangle \otimes 
\ldots \otimes |\eta\rangle$, so that measurements (projections) of any one part do not affect the states of any other part. Determining whether or not a mixed state $\rho = \sum_i p_i |\Psi_i\rangle \langle \Psi_i|$ is unentangled is more difficult because the decomposition of $\rho$ is not unique:  $\rho$ is unentangled if there is a decomposition for which each term corresponds to an unentangled pure state.  

The simplest system that can be entangled is a two-dimensional bipartite state of the general form
$|\Psi\rangle = a|++\rangle + b|+-\rangle + c|-+|\rangle + d|--\rangle$, which is factorizable 
only if the quantity 
$e = |ad-bc| = 0$.  The largest possible value of $e$ is $e = 1/2$, and this is what is obtained  for maximally entangled states, such as any of the four Bell states (for example $(|++\rangle + |--\rangle)/\sqrt 2$).  All quantitative formulations of entanglement on this simple system, including entanglement entropy and concurrence, reduce to 
monotonically increasing functions of $e$. 
However, for larger systems there are several different formulations that generalize this calculation, corresponding to all the different ways a multi-partite system can be partitioned, and how the various parts are entangled with each other. Regardless of which formulation is chosen, if $C(\Psi_i)$ represents the entanglement of a pure state $|\Psi_i\rangle$, the corresponding entanglement of a density is $$C(\rho) = 
\inf_{\{p_i,  \Psi_i\}} \sum_i p_i C(\Psi_i),$$  where the {\em infimum} selects the decomposition of $\rho$ 
that minimizes $\sum_i p_i C(\Psi_i)$ \cite{mintert2005,UhlmannPhysRevA.62.032307}. 

In this paper we will consider densities that are thermal mixtures, $\rho = \frac{1}{Z} \sum_i e^{-E_i/T}|\Psi_i\rangle \langle \Psi_i|$, where $T$ is the temperature, 
$|\Psi_i\rangle$ are the eigenstates of a Hamiltonian $H$ with energy eigenvalues $E_i$, and $Z$ is the partition function, $Z = \sum_i e^{-E_i/T}$.  
In highly symmetric systems the energy eigenstates will be superpositions of states related by symmetry and are often highly entangled as result.  However, symmetry also produces degenerate multiplets, which, under some circumstances, leads to a decomposition of the density into terms corresponding to pure states that are less entangled than the original eigenstates of the Hamiltonian.  In other words, symmetry tends to introduce entangled eigenstates of the Hamiltonian while at the same time symmetry-required degeneracies may reduce entanglement of the thermal mixed state. This paper investigates the competition between these effects.
This paper also introduces a new approach to evaluating 
the entanglement of a highly symmetric mixed state by explicitly constructing unentangled contributions to the density that are also invariant under all of the symmetries of the Hamiltonian.  Using these constructions we greatly simplify the problem of evaluating the infimum. 

We begin by reviewing entanglement measures for pure states and  their extension to mixed states.  In Section 3, we develop a method to quantify entanglement in symmetric spin systems which makes use of explicitly constructed unentangled density matrices that are invariant under the symmetry group of the Hamiltonian. Finally, we apply our method to three-spin systems to study entanglement as a function of the Hamiltonian parameters (coupling constants) and temperatute. 


\section{Entanglement and Concurrence}
\subsection{Concurrence  measures for pure states}

For the general bipartite two-dimensional state, $\lvert \Psi\rangle = a|++\rangle + b|+-\rangle + c|-+\rangle + d|--\rangle$, the concurrence is given by 
\begin{equation}
    C(\Psi) = |\langle \tilde{\Psi}|\Psi\rangle|
\end{equation}
where $|\tilde{\Psi}\rangle$ is the `spin-flipped' $|\Psi\rangle$,
$|\tilde{\Psi}\rangle =
\sigma_y\otimes \sigma_y
|\Psi^{*}\rangle$, where $\sigma_y = \left(\begin{array}{cc}
0 & -i \\ i & 0 \end{array}\right)$ and
$|\Psi^{*}\rangle = 
a^{*}|++\rangle + b^{*}|+-\rangle + c^{*}|-+\rangle + d^{*}|--\rangle$. (If the states $|\pm\rangle$ actually refer to spin then $|\tilde{\Psi}\rangle$ is the time-reversal of $|\Psi\rangle$.)  This yields 
\begin{equation}
C( \Psi) = 
2|ad - bc|,
\label{concur-bipart}
\end{equation}
which ranges from 0, when the state is separable, to 1, when it is maximally entangled. 
  There are other, different ways to express the concurrence of a bipartite state, leading to different  generalizations of the concurrence for multi-partite states,  which can illuminate different aspects of entanglement, for example, the entanglement of different parts with each other, or the entanglement within a part. 
  One such expression is
\begin{equation}
  C(\Psi) = \sqrt{2 - \sum_i \mbox{Tr} \rho_i^2},
  \label{eq:bipart-concur}
  \end{equation}
 where $\rho_{1,2}$ are the reduced densities obtained by taking the traces over each of the two parts of the density $\rho = |\Psi\rangle\langle \Psi|$. In general, the trace of $\rho_i^2$ is equal to the trace of  $\rho_{\bar{i}}^2$ when $i$ and $\bar{i}$ are complementary parts. Here we find that
 $\mbox{Tr} \rho_1^2 = \mbox{Tr} \rho_2^2 
 = 1-2|ad-bc|^2$, and so $C(\Psi) = 2|ad-bc|$ as required.  Note that the maximum value of $\mbox{Tr} \rho_i^2$ is one, and this corresponds to an unentangled state, so that the minimum of $C(\Psi)$ is zero.  The minimum value of  $\mbox{Tr} \rho_i^2$ is $1/2$, corresponding to a maximally entangled state, with $C(\Psi) = 1$ in this case.

 The expression (\ref{eq:bipart-concur}) of bipartite concurrence is easily generalized to an $N$-partite state $|\Psi\rangle$ \cite{carvelloPhysRevLett.93.230501},
\begin{equation}
C( \Psi) = 2^{1 - \tfrac{N}{2}} \sqrt{(2^N - 2) - \sum_i \operatorname{Tr}(\rho_i^2)},
\label{N-concur}
\end{equation}
where $\rho_i$ are the reduced densities of 
$\rho = |\Psi\rangle \langle \Psi|$ resulting from all possible ways of partitioning the system into two parts.  For an $N$-partite state, there will be a total of $2^N -2$ terms in the sum over $i$. If the density is unentangled then $\mbox{Tr} \rho_i^2 = 1$ for all of the  reduced densities $\rho_i$, and the concurrence vanishes.  This concurrence measure is useful for our purposes because it 
is invariant under permutations of any of the parts.  While the formulation (\ref{N-concur}) can be applied states with parts that have more than two levels (two dimensions), in the following we will only consider two-level systems, such as spins or qubits. 

\subsection{Concurrence of mixed states}

The amount of entanglement of mixed states is generally much more  difficult to evaluate than the concurrence of pure states because there are infinitely many decompositions of the density to consider.
In general, a density
$\rho = \sum_i p_i |\Psi_i\rangle \langle \Psi_i|$, where $|\Psi_i\rangle$ are normalized pure states, $p_i \geq 0$ and $\sum_ip_i = 1$, can be re-expressed in terms of a different set of normalized pure states 
$|\Phi_i\rangle$, as
\begin{equation}
    \rho  =    \sum_i^n p_i |\Psi_i\rangle \langle \Psi_i| =  \sum_j^m q_j|\Phi_j\rangle \langle \Phi_j| 
    \end{equation}
  where $q_i \geq 0$,
    and the concurrence of $\rho$ is obtained by finding the set of states 
    $|\Phi_j\rangle$ that minimizes
    $\sum_j q_j C(\Phi_j)$, that is, the {\em convex roof} \cite{UhlmannPhysRevA.62.032307}
\begin{equation}
        C(\rho) =  \inf_{\{q_i,  \Phi_i\}}\sum_j q_j C(\Phi_j).
    \end{equation}  Here $C(\Phi)$ can be any measure of concurrence for a pure state $|\Phi\rangle$; in the following we will use the measure given by Eq.\ \ref{N-concur}.
Note that while the set of states $|\Psi_i\rangle$ will often be a complete orthonormal set (for example, the eigenstates of a Hamiltonian),  the states $|\Phi_j\rangle$ need not be, and there is no limit to the number of terms in the sum over $j$.

One can re-write the density as \cite{mintert2005, hughston1993} \begin{equation}
   \rho  =  \sum_i^n |\psi_i\rangle \langle \psi_i| 
    =  \sum_j^m |\phi_j\rangle \langle \phi_j| 
\end{equation}
where $
\ket{\psi_i} = \sqrt{p_i}\ket{\Psi_i}
$ and $|\phi_j\rangle = \sqrt{q_j}|\Phi_j\rangle$. The sets of states $|\psi_i\rangle$ and 
$|\phi_j\rangle$ are related by a left unitary transformation, 
$$
|\psi_i\rangle =
\sum_j U_{ij} |\psi_j\rangle,$$
where $U$ is an
$n\times m$ {\em left-unitary} matrix with $n \leq m$ and
$\sum_i^n U_{ki}^{\dagger} U_{ij}
=\delta_{kj}$.  Evaluating the infimum amounts to a minimization over all left-unitary matrices with all numbers of columns $m$.
However, the minimization problem is avoided in bipartite, two-dimensional systems where there is an exact algebraic formula  which uses the spin-flipped density matrix $\tilde{\rho} = (\sigma_y \otimes \sigma_y)\rho^*(\sigma_y \otimes \sigma_y)$.  The concurrence is given by    \cite{Hill-WootersPhysRevLett.78.5022,WootersPhysRevLett.80.2245}
\begin{equation}
C(\rho) = \max(0, \lambda_1 - \lambda_2 - \lambda_3 - \lambda_4)
\label{wootter}
\end{equation}
where $\lambda_i$ are the ordered (largest to smallest) eigenvalues of the matrix
$$
R = \sqrt{\rho^{1/2} \tilde{\rho}\rho^{1/2}}.
$$
 


\section{Symmetry-Guided Decomposition of the Density Matrix}
Our approach to quantifying entanglement in symmetric systems is to explicitly decompose the mixed state into a sum of an unentangled part and (if necessary) an entangled remainder. The goal is to find a decomposition 
that approaches the infimum in the convex roof definition and that produces a smooth, monotonically increasing result as a function of the model parameters between endpoints that are known to be unentangled and maximally entangled states. 
The procedure can be outlined as follows:
\begin{enumerate}
\item 
Define a set of axes that reflects the symmetry of the system.  The axis of highest rotational symmetry is selected as the global $z$-axis.  Global $x$ and $y$ axes should also be symmetry axes, if possible, and the three axes should obey the right-hand rule.  For spin systems, basis states are normally written as $|\pm\pm\ldots \pm\rangle$  where the order of the entries corresponds to an ordered set of positions (sites) where the spins reside. 
For each position at which the spins reside define a set of local axes, where the local $z$ axis is the axis of highest rotational symmetry of the site. The local axes may be different for each site, and for each site the spin quantization axis will be the local $z$-axis. Thus a basis state of the form 
$|+++\rangle$ is a state where the three spins are pointing `up' in the direction of their local $z$-axes, and
does not imply that all three spins point in the same direction.  States expressed with respect to any other axes
will be indicated with subscripts, for example
$|\pm\rangle_x = (|+\rangle \pm |-\rangle)/\sqrt{2}$ and $|\pm\rangle_y = (|+\rangle \pm i|-\rangle)/\sqrt{2}$.

\item Construct and solve the Hamiltonian $H$ that reflects the symmetries of the system. Since we will be dealing with symmetric arrangements of interacting spins, the symmetry group will include rotations (continuous or discrete), improper rotations (reflections), and time reversal. 
Generally, the Hamiltonian can be block-diagonalized where different blocks correspond to sets of {\em symmetrized} basis states belonging to different irreducible representations (IR's) of the symmetry group.  We find these states explicitly and compute their concurrence using Eq.\ \ref{N-concur}.  The eigenstates of $H$ will be linear combinations of these basis states, and their concurrence can also be calculated. 
Note that IR's that are multi-dimensional will produce sets of degenerate eigenstates of the Hamiltonian.  Lastly we construct the thermal density using the eigenstates of $H$. The thermal density will be invariant under the symmetry group, as well as the separate contributions from each set of degenerate eigenstates. We then express $\rho$ such that the degenerate eigenstates are grouped together by
writing $\rho = \sum_i p_i  \rho_i$, where $p_i = d_i e^{-E_i/T}/Z$, $d_i$ is the number of degenerate states with eigenvalues $E_i$, and $\rho_i = \frac{1}{d_i} \sum_{j=1}^{d_i}|\Psi_{i,j}\rangle\langle \Psi_{i,j}|$.

\item Determine the independent components of the density matrix that are allowed by symmetry.  For example, in a three spin system, $\rho_{+++,+++} = \rho_{---,---}$ because of time reversal symmetry.   Beginning with the diagonal elements, arrange all of the independent components of the  density in a one-dimensional array.  The thermal density $\rho$ and each of its individual terms $\rho_i$ can be represented in this way, as well as any mixed state that is invariant under the symmetry group.

\item Generate a set of separable, symmetric  mixed states, $\eta_k$. These states will be built by considering all possible spin states expressed with respect to the local $x$, $y$ and $z$ axes. 
  \item Find a new decomposition of the thermal density $\rho$ that optimally includes the separable, symmetric  mixed states found in Step 4. If the original decomposition of the thermal density is
  $\rho = \sum_i p_i \rho_i$, after optimization the thermal density is decomposed as
  $\rho = \sum_i p'_i \rho_i + \sum_k q_k \eta_k$.
  The optimization is performed such that the quantity $C_s(\rho)  \equiv \sum_{i}  \frac{1}{d_i} p'_i C(\Psi_i)$, where $d_i$ is the degeneracy of the state $|\Psi_i\rangle$, is minimized, subject to the constraints that
  $0 \leq p_i', q_k \leq 1$ and $\sum_i p'_i + \sum_k q_k = 1$.
  We \underline{define}
  the quantity $C_s(\rho)$ obtained by this procedure  to be the concurrence of the state $\rho$. 
  \end{enumerate}
  \subsection{Examples}
\begin{center}
\begin{figure}[h]
\begin{minipage}{0.4\textwidth}
\includegraphics[width=\linewidth]{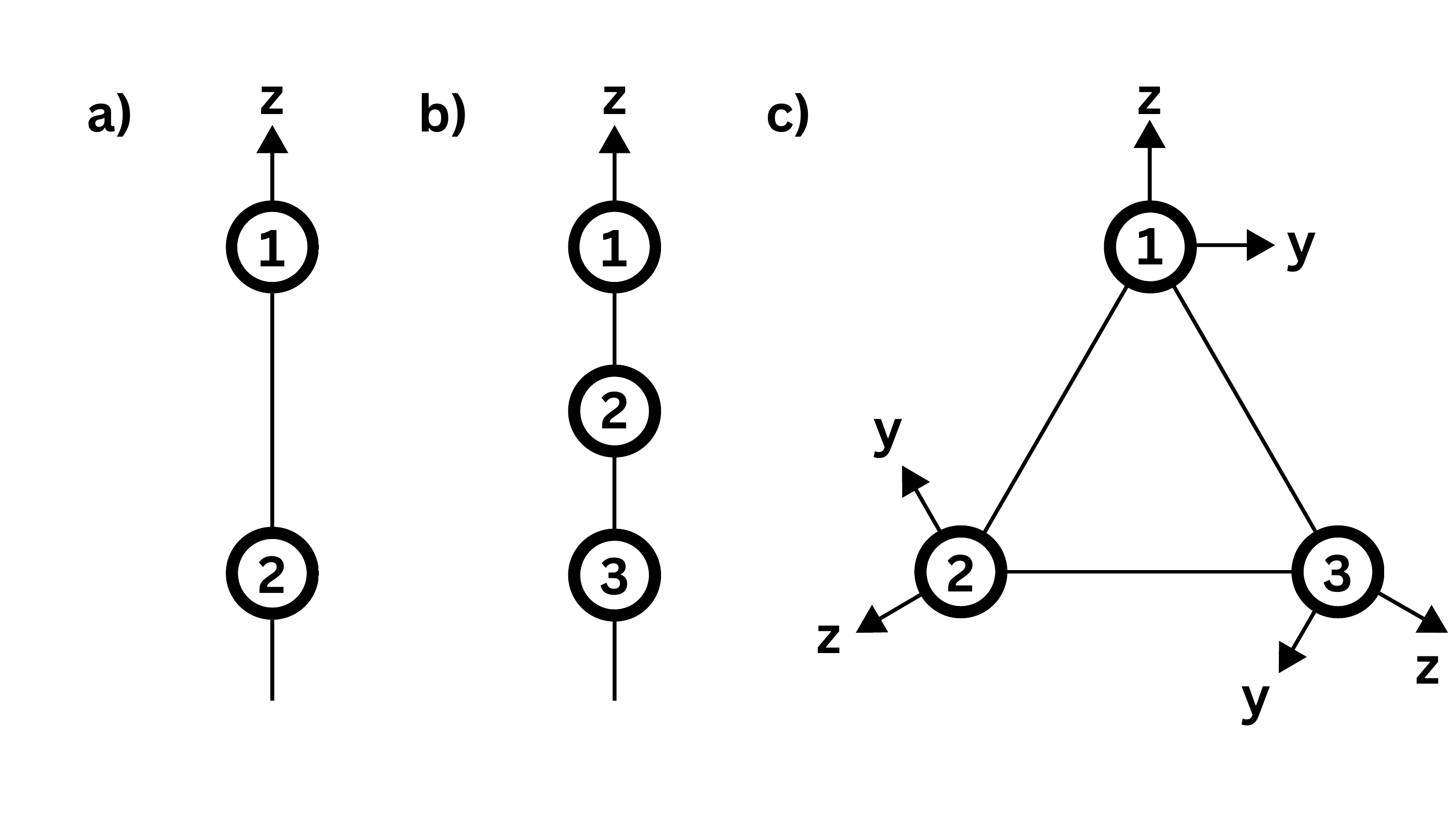}
\end{minipage}
\caption{a): Two spins on a line; b): Three spins on a line; c): three spins arranged on a triangle. The arrows and labels indicate the local axes that will be used for the spin states.\label{fi:spin_arrange}}
\end{figure}
\end{center}

In the following we  apply our method to four examples, beginning with the simplest possible case of two interacting spins with full rotational symmetry. Our other examples will discuss spins arranged as shown in Fig.\ \ref{fi:spin_arrange}.
\subsubsection{Two interacting spins: isotropic}
  The eigenstates of two interacting spins in an isotropic environment will be the singlet and triplet states,
  $$|s\rangle = 
  (|+-\rangle - |-+\rangle)/\sqrt{2}$$
  and 
  \begin{eqnarray}
|t_{+1}\rangle &=& |++\rangle \nonumber \\
|t_0\rangle & = & (|+-\rangle + |-+\rangle)/\sqrt{2} \nonumber \\
|t_{-1}\rangle & = & |--\rangle.
\nonumber
\end{eqnarray}
These are the eigenstates of $H = 
{\cal J} S_1\cdot S_2$, which is the only possible isotropic spin-spin interaction, with eigenvalues $-3{\cal J}/4$ and ${\cal J}/4$. 
The associated densities $\rho_s =|s\rangle\langle s|$ and 
$\rho_t = 
\frac{1}{3}\sum_{i}|t_i\rangle\langle t_i|$. 
  The concurrences of these states are 
  $C(s) = C(t_0) = 1$ and $C(t_{\pm1}) =0$.  However,  $C(\rho_t) = 0$ because it is  equivalent to the separable state $\eta_1$ (see Eq. \ref{rho_t}, below), while $C(\rho_s) = 1$.


We construct two unentangled symmetric densities,
\begin{eqnarray}
\eta_1 & = & \frac{1}{6}
\sum_{\alpha = x,y,z}
|++\rangle\langle ++|_{\alpha}+ |--\rangle\langle --|_{\alpha} \nonumber \\
& = & \rho_t
\label{rho_t}\\
\eta_2 & = & \frac{1}{6}
\sum_{\alpha = x,y,z}
|+-\rangle\langle +-|_{\alpha}+ |-+\rangle\langle -+|_{\alpha}\nonumber \\
& = & \rho_s/2 + \rho_t/2,
\end{eqnarray}
where their expressions in terms of $\rho_s$ and $\rho_t$ can easily be verified.
Thus the thermal state
$$
\rho = p_s\rho_s + p_t \rho_t,
$$
can be optimally decomposed as
$$
\rho = \left\{
\begin{array}{ll}
(p_t-p_s)\eta_1 + 2 p_s \eta_2 & p_t > p_s\\
(p_s-p_t)\rho_s + 2p_t \eta_2 & p_s > p_t.
\end{array} 
\right. 
$$
Therefore the concurrence is
$$
C_s(\rho) = \left\{
\begin{array}{ll}
0 & p_t > p_s\\
p_s-p_t & p_s > p_t.
\end{array} 
\right. 
$$
It is straightforward to verify that this result is equal to what is obtained using Wootter's formula Eq.\ \ref{wootter}:  here $\tilde{\rho} = \rho$, and so $R= \rho$, and the eigenvalues of $\rho$ are $p_s$ (non-degenerate) and $p_t/3$ (triply degenerate). 

\subsubsection{Two interacting spins on a symmetry axis}

Instead of a full three-dimensional rotation group, the symmetry elements are continuous rotations about the $z$-axis, as well discrete rotations and reflections in other directions.  We also have time reversal symmetry, but it is redundant in this case. The local axes will be the same as the global axes, as shown in Fig.~1. The most general Hamiltonian is 
$$
H = {\cal J}_1 S_{1z}S_{2z} + {\cal J}_2(S_{1x}S_{2x} + S_{1y}S_{2y}),
$$
where ${\cal J}_{1,2}$ are coupling constants and $S_{i\alpha}$ are spin operators.
The total spin $S$ and $S_z$ are conserved, and so we have non-degenerate eigenstates $|s\rangle$ and $|t_0\rangle$, while
$|t_{\pm1}\rangle$ are degenerate, with eigenvalues $E_s =-{\cal J}_1/4-{\cal J}_2/2$, $E_{t_0}= -{\cal J}_1/4 + {\cal J}_2/2$ and $E_{t_{\pm1}} = {\cal J}_1/4$.  The associated densities are 
$\rho_s$, 
$\rho_{t_{\pm1}} = \frac{1}{2}(|t_{+1}\rangle\langle t_{+1}|+ |t_{-1}\rangle\langle t_{-1}|)$ and
$\rho_{t_0} = |t_0\rangle\langle t_0|$, and the thermal density is 
\begin{equation}
\rho = p_s\rho_s + p_{t_0}\rho_{t_0} + p_{t_{\pm1}} \rho_{t_{\pm1}},
\label{eq:2thermal}
\end{equation}
where 
$p_s = e^{-E_s/T}/Z$, 
$p_{t_0} =e^{-E_s/T}/Z$ and $p_{t_{\pm1}} = 2e^{-E_{t_{\pm1}}/T}/Z$. Note that $\rho_s$ and $\rho_{t_0}$ are the only entangled contributions; our optimal decomposition will decrease their coefficients as much as possible.

In the previous example, the general symmetric density matrix had only two independent elements; here there are three real elements,
  $$\rho = \left(\begin{array}{cccc}
  a/2 \\
  & b/2 & c \\
  & c & b/2 \\
  & & & a/2\end{array}\right)\equiv[a,b,c].$$
 Using the same notation, the terms in the thermal density are:
\begin{eqnarray}
\rho_s &= &[0,1, -1/2]\\
\rho_{t_{\pm1}} &= &[1,0,0]\\
\rho_{t_0} & = & [0,1,1/2].
\end{eqnarray}

According to our method, we consider all 36 unentangled states of the form $|\pm\pm\rangle_{\alpha \beta}$ 
($\equiv |+\rangle_{\alpha}\otimes |\pm\rangle_{\beta}\rangle$, 
$\alpha, \beta = x,y,z)$, but
most of these will not appear in symmetric densities. Altogether there are four different (but not independent) symmetric densities, 
\begin{eqnarray}
\eta_{1} & = & \frac{1}{2}
(|++\rangle\langle ++|+ |--\rangle\langle --|) \nonumber \\
& = & [1,0,0]= \rho_{t_{\pm1}}
\\
\eta_{2}& = & \frac{1}{4}
\sum_{\alpha = x,y}
|++\rangle\langle ++|_{\alpha} + |--\rangle\langle --|_{\alpha}\nonumber  \\
& = & [1/2,1/2,1/4]
= \rho_{t_0}/2 + \rho_{t_{\pm1}}/2\\
\eta_3 & = & \frac{1}{2}
(|+-\rangle\langle +-| +
|-+\rangle\langle -+|)\\
& = & [0,1,0]
= \rho_s/2 + \rho_{t_0}/2 \nonumber \\
\eta_4 & = & \frac{1}{3}
\sum_{\alpha = x,y}
|+-\rangle\langle +-|_{\alpha}+ |-+\rangle\langle -+|_{\alpha}\nonumber \\
& = & [1/2,1/2,-1/4] = \rho_s/2 + \rho_{t_{\pm1}}/2
\end{eqnarray}
which we include in a new decomposition of the thermal density,
$$
    \rho = p_s'\rho_s +
    p_{t_0}'\rho_{t_0}
    + p'_{t_{\pm1}}\eta_1
    + q_2\eta_2 + q_3 \eta_3 + q_4 \eta_4
$$
with $p_s' = p_s -\frac{q_3}{2} - \frac{ q_4}{2}, $
$p_{t_0}' = p_{t_0} -\frac{q_2}{2} -\frac{q_3}{2}$ and
$p'_{t_{\pm1}} = p_{t_{\pm1}} -\frac{q_2}{2} - \frac{q_4}{2}$. The  concurrence thus obtained will be 
$$
C_s(\rho) = p_s' + p_{t_0}'.
$$
The exact form that the  optimal decomposition takes will depend on the original parameters $p_i$.  In general,  $p_s'$ and $p_t'$ will vanish if $p_s < 1/2$ and $p_t < 1/2$, yielding $C_s(\rho) = 0$.  The full result is
$$
C_s(\rho) = \left\{
\begin{array}{cc}
2 p_s - 1 & p_s >1/2 \\
2 p_{t_0} -1 & p_{t_0} > 1/2 \\
0 & \mbox{ otherwise}
\end{array} 
\right. ,
$$
in agreement with Wootter's formula Eq.\ \ref{wootter}.

\begin{figure}[ht]
\centering
\begin{minipage}{0.4\textwidth}
\includegraphics[width=\linewidth]{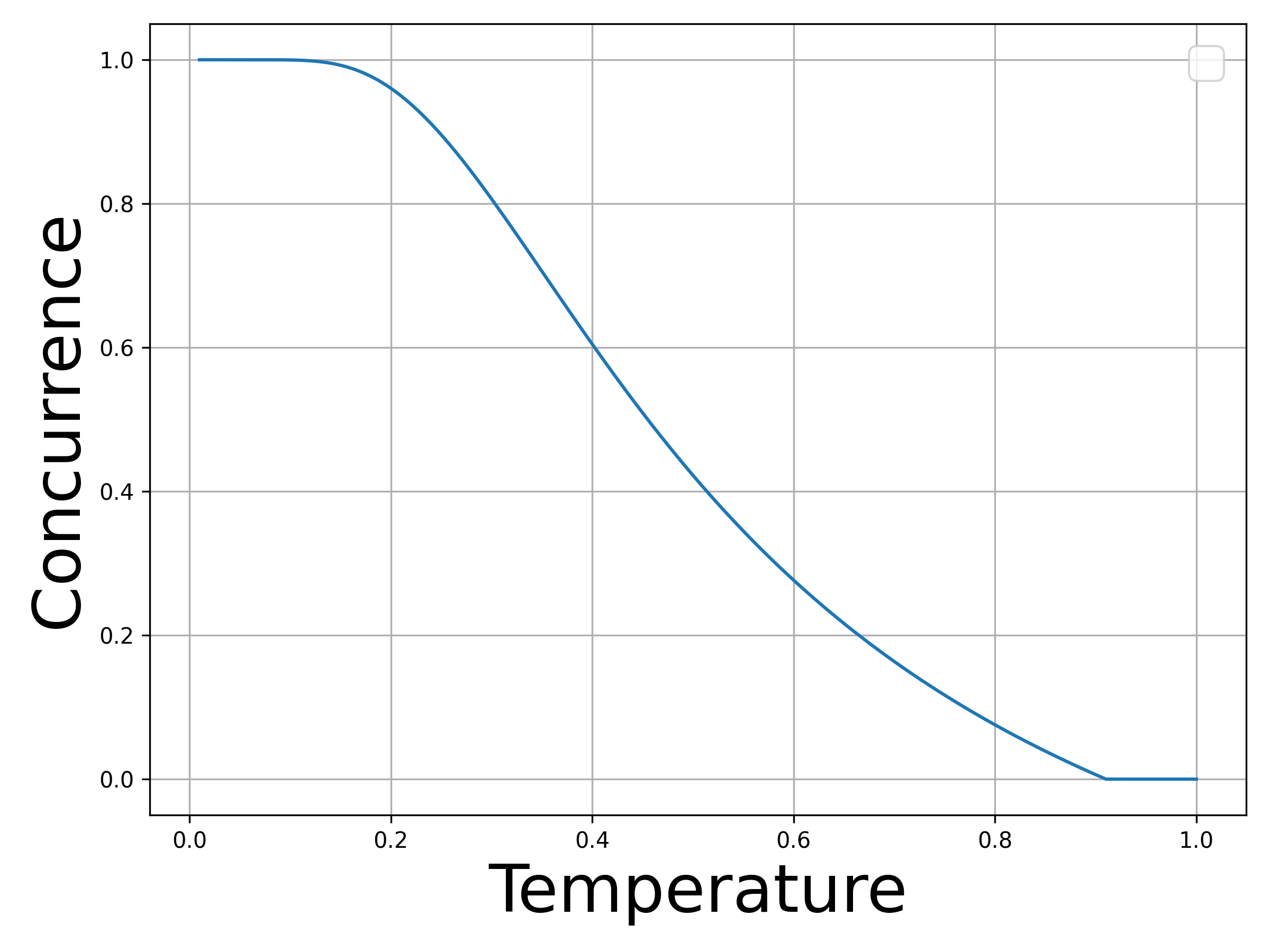}
\end{minipage}
\caption{Concurrence $C_s$ as a function of temperature for a two-spin system with ${\cal J}_1 = {\cal J}_2 =1$.
\label{fi:2pinT}}
\end{figure}

It is interesting to note that while $C_s(\rho)$ is a continuous function of the parameters $p_s$ and $p_{t_0}$, it is not smooth.  Therefore the concurrence of  a thermal density (\ref{eq:2thermal})  is not a smooth function of temperature either.  Fig.\ \ref{fi:2pinT} 
shows the concurrence of the 2-spin system as a function of temperature for ${\cal J}_1 = {\cal J}_2 =1$. 
The singlet state is the ground state, and as $T \rightarrow 0$ $p_s \rightarrow 1$, while for large $T$, $p_s, p_{t_0} \rightarrow 1/4$ and the concurrence vanishes. There is a discontinuity in the slope at $T \approx 0.91$, where $p_s = 1/2$.


\subsubsection{Three spins on a line}
Here we consider three equally spaced spins arranged on a line, which has the same symmetry group as two spins. The axis for continuous rotations is selected as the global $z$-axis and the local $z$-axis for all three spins.  

There are 8 symmetrized basis states, which can be sorted into time-reversed pairs:
\\
$|\Psi_{1\pm}\rangle = \{|+++\rangle, |---\rangle$\}\\
$|\Psi_{2\pm}\rangle = \{\frac{|--+\rangle-|+--\rangle}{\sqrt{2}},\frac{|++-\rangle-|-++\rangle}{\sqrt{2}}\}$\\
$|\Psi^{(1)}_{3\pm}\rangle = \{|-+-\rangle, |+-+\rangle\}$\\
$|\Psi^{(2)}_{3\pm}\rangle = 
\{
 \frac{|--+\rangle+|+--\rangle}{\sqrt{2}},|
 \frac{|++-\rangle+|-++\rangle}{\sqrt{2}}\}$\\
 where the lower index indicates different irreducible representations of the symmetry group to which the states belong.  The states are distinguished by how they transform under the symmetry operations; a representative set is shown in the table below, where ${\cal K}$ stands for time reversal, $C_{2y}$ is a 
 rotation of $\pi$  about the $y$-axis, $I$ is inversion (about the central site), and $R_z(\theta)$ is a rotation about the $z$ axis by an angle $\theta$:\\
 \begin{tabular}{c|cccc}
 state &  ${\cal K}$ & $C_{2y}$ & $I$ & $R_z(\theta)$ \\
 \hline
$ |\Psi_{1\pm}\rangle$ & $\pm|\Psi_{1\mp}\rangle$ & $\pm|\Psi_{1\mp}\rangle$ &$|\Psi_{1\pm}\rangle$ & $e^{\mp i 3\theta/2}|\Psi_{1\pm}\rangle$ \\
$|\Psi_{2\pm} \rangle$ & $\pm|\Psi_{2\mp}\rangle$ & $\mp|\Psi_{2\mp}\rangle$
&$-|\Psi_{2\pm}\rangle$ & $e^{\pm i \theta/2}|\Psi_{2\pm}\rangle$ \\
$ |\Psi_{3\pm}\rangle$ & $\pm|\Psi_{3\mp}\rangle$ & $\pm|\Psi_{3\mp}\rangle$
& $|\Psi_{3\pm}\rangle$ & $e^{\pm i \theta/2}|\Psi_{3\pm}\rangle$ 
 \end{tabular}
 
 \vspace{.05in}
\noindent 
Since they are 1/2-integral, all of the states acquire a phase of $-1$ for $2\pi$ rotations. 
The states $|\Psi^{(1)}_3\rangle $ and  $|\Psi^{(2)}_3\rangle$ belong to the same irreducible representation, so we expect them to be mixed by the Hamiltonian.

The most general symmetry-invariant spin-spin Hamiltonian is
\begin{eqnarray}
H & =&  {\cal J}_1 (S_{1z} S_{2z}+S_{2z} S_{3z}) \nonumber \\
& + & {\cal J}_2(S_{1x}S_{2x}+S_{1y}S_{2y} + S_{2x}S_{3x} + S_{2y}S_{3y})\nonumber  \\
&+ & 
   {\cal J}_3 S_{1z} S_{3z}+{\cal J}_4 (S_{1x}S_{3x} + S_{1y}S_{3y}) 
\end{eqnarray}
In the basis of the symmetrized states $|\Psi_{+}\rangle$, the Hamiltonian is 
$$H = \left(\begin{array}{cccc}
\frac{{\cal J}_1}{2} + \frac{{\cal J}_3}{4} \\
& -\frac{{\cal J}_3}{4} -\frac{{\cal J}_4}{2} \\
& & -\frac{{\cal J}_1}{2} + \frac{{\cal J}_3}{4} & \frac{{\cal J}_2}{\sqrt 2} \\
& & \frac{{\cal J}_2}{\sqrt 2} & -\frac{{\cal J}_3}{4} +\frac{{\cal J}_4}{2}
\end{array} \right)$$ 
with an identical matrix for the $|\Psi_{-}\rangle$ states.  The $2\times 2$ block can be solved to find the eigenvectors
$|\Psi_{3,4}\rangle = 
 \alpha |\Psi_{3}^{(1,2)}\rangle \pm \beta|\Psi_3^{(2,1)}\rangle$, where
$\alpha$ and $\beta$ are real, with associated eigenvalues $E_{3,4}$. Lastly, we denote the concurrences of the eigenstates, calculated using Eq. \ref{N-concur}, as $C_i = C(\Psi_i)$, where
\begin{eqnarray}
C_1 & = & 0 \\
C_2 & = & 1 \\
C_3  & = & \sqrt{3-(3\alpha^4 + 2\alpha^2\beta^2 + 2\beta^4)} \\
C_4 & = & \sqrt{3-(3\beta^4 + 2\alpha^2\beta^2 + 2\alpha^4)}
\label{eq:C4}
\end{eqnarray}


The general form of the density matrix is
$$
\left(
\begin{array}{rrrrrrrr}
a/2 &  &&  & &&  \\ 
& b/4 &  d& e &  &  \\
& d& c/2 & d &  & & \\
& e & d& b/4 & &  & \\
 &  &  &&b/4&d &e & \\
 & & & &d&c/2&d & \\
 & & & & e&d&b/4& \\
 & &  & & & & & a/2 \end{array}\right)
$$
$$
\equiv[a,b,c,d,e],$$
where the order of the elements in the density matrix is given by $\{ |+++\rangle, |++-\rangle, |+-+\rangle, |-++\rangle, |+--\rangle, |-+-\rangle, |--+\rangle, |---\rangle\}$ so as to make the symmetries of $\rho$ more evident.


There are four symmetric terms in the thermal density matrix, corresponding to the four eigenvalues of $H$, 
$$\rho_i = \frac{1}{2}(|\Psi_{i+}\rangle\langle \Psi_{i+}| + |\Psi_{i-}\rangle\langle \Psi_{i-}|
).$$  
Using the notation defined above, these are 
\begin{eqnarray}
    \rho_1 & = & [1,0,0,0,0] \\
    \rho_2 & = & [0,1,0, 0,-1/4] \\
    \rho_3 & = & [0, \beta^2, \alpha^2, \alpha\beta/2\sqrt{2}, \beta^2/4]\\
    \rho_4 & = & 
[0, \alpha^2, \beta^2, -\alpha\beta/2\sqrt{2}, \alpha^2/4]    
\end{eqnarray}

In three-spin systems, we will consider each of $6^3= 216$ separable states can be arranged into separable symmetric densities, for example, 
$$
   \eta_1 =\frac{1}{2}(|+++\rangle\langle +++| + |---\rangle \langle ---|).
$$
The results are summarized here:\\

\vspace{0.05in}
\begin{tabular}{ll}
    $\eta_1  = [1,0,0,0,0]$ & $|+++\rangle$ \\
$    \eta_2  =  [0,1,0,0,0]$ &$|++-\rangle$\\
$    \eta_3 = [0,0,1,0,0]$ & $|+-+\rangle$ \\
$\eta_4  = [1/4,1/2,1/4,1/16,0]$ & $|+++\rangle_{zxx}$ \\
$\eta_5 =  [1/4,1/2,1/4,-1/16,0]$ &
$|++-\rangle_{zxx}$ \\
$\eta_6  =  [1/4,1/2,1/4,0,-1/8] $ &
$|++-\rangle_{xzx}$\\
$\eta_7 = [1/4,1/2,1/4,0,1/8]$ &$|+++\rangle_{zxz}$\\
$\eta_8 = [1/4,1/2,1/4,1/8,1/8]$ & $|+++\rangle_{xxx}$\\
$\eta_9  = [1/4,1/2,1/4,-1/8,1/8]$ &$|+-+\rangle_{xxx}$
\end{tabular}\\
\vspace{0.05in}

\noindent where those symmetric densities that can be expressed as  a positive sum of other symmetric densities have been omitted, as these will be redundant in our optimization procedure. The second column lists examples of separable states that are used to build the separable symmetric densities, where omitted subscripts indicates that all three spins are expressed with respect to the $z$ axis.

The thermal density 
is
$$
\rho = \sum_{i=1}^4 p_i\rho_i
$$
where $p_i = e^{-E_i/T}/Z$.
The optimized form of $\rho$ is
$$
\rho = \sum_{i=1}^4 p_i' \rho_i
+ \sum_{j=1}^9 q_j\eta_j $$
and the concurrence thus obtained will be
$$
C_s(\rho) = \sum_i p_i' C_i.
$$
Our results for this system are presented in Section 5.1.

\subsection{Three Spins Arranged in a Triangle}
\label{sec:triangle3spins}
In this case the spins are arranged as shown in Fig.\ \ref{fi:spin_arrange}c). The global $z$-axis points out of the page, the global $x$ axis points horizontally to the right, and the global $y$ axis points vertically. 
The local $y$ and $z$ axes of each site are shown in the figure; each  local $x$-axis points out of the page (in the direction of the global $z$ axis) for all three sites.  The local axes for each site expressed in terms of the global axes is:
\begin{eqnarray}
 \hat{x}_1 & = &  (0,0,1) \equiv \hat{z} \\
\hat{y}_1 & =& (1,0,0) \equiv  \hat{x} \\
\hat{z}_1&  = &(0,1,0) \equiv \hat{y}\\
\hat{x}_2 & = &(0,0,1) \equiv \hat{z} \\
\hat{y}_2 & = & ( -1/2, +\sqrt 3/2,0) \equiv \hat{x}' \\
\hat{z}_2 &= &(-\sqrt{3}/2, -1/2,0) \equiv \hat{y}'\\
\hat{x}_3 &=& (0,0,1) \equiv \hat{z} \\
\hat{y}_3 &=& ( -1/2, -\sqrt 3/2,0) \equiv \hat{x}''   \\
\hat{z}_3&  =& (\sqrt{3}/2, -1/2,0)  \equiv \hat{y}''
\end{eqnarray}
The Hamiltonian, its eigenstates, and the set of unentangled symmetrized states, and all densities  will be expressed with respect to the  local axes.

The symmetry group is $D_{3h}'$, which includes rotations  about global axes, $C_{3z}$ ($2\pi/3$), $(C_{3z})^2$ ($4\pi/3$), $C_{2y}$, $C_{2y'}$ and $C_{2y''}$, and reflections through the planes perpendicular to the rotation axes.  Since the total spin is 1/2-integral, the symmetry group includes an equal number of additional rotations that are obtained by adding $2\pi$ rad to the rotations given above. Time-reversal symmetry ${\cal K}$ is also present.

There 8 symmetrized basis states are:
\begin{eqnarray}
    |\Psi^{(1)}_{1\pm}\rangle 
   &  = & \{|+++\rangle, |---\rangle\} \nonumber \\
    |\Psi^{(2)}_{1\pm} \rangle 
    & = & \left\{\frac{1}{\sqrt 3}(|+--\rangle + |-+-\rangle + |-++\rangle), \right. \nonumber  \\
& &     \left.   \frac{1}{\sqrt 3} (|-++\rangle + |+-+\rangle + |++-\rangle)\right\} \nonumber \\
|\Psi_{3\pm}\rangle & =   &
\frac{1}{\sqrt 2} \left(
|\Phi_{a\pm}\rangle \mp |\Phi_{b\mp}\rangle \right) \\
|\Psi_{4\pm}\rangle & = &  
\frac{1}{\sqrt 2} \left(
|\Phi_{b\pm}\rangle \mp |\Phi_{a\mp}\rangle \right) 
\end{eqnarray}
where
\begin{align*}
     |\Phi_a\pm\rangle   = \left\{\frac{1}{\sqrt 3}(|--+\rangle + \epsilon |-+-\rangle + \epsilon^2|+--\rangle), \right. \\
 \left.    \frac{1}{\sqrt 3}(|++-\rangle + \epsilon^2|+-+\rangle + \epsilon |-++\rangle)\right\} \\
    \ket{\Phi_{b\pm}} = 
    \left\{
    \frac{1}{\sqrt 3} 
    (|--+\rangle + \epsilon^2 |-+-\rangle + \epsilon|+--\rangle) \right. \\
    \left. 
    \frac{1}{\sqrt 3}
    (|++-\rangle + \epsilon|+-+\rangle + \epsilon^2 |-++\rangle)\right\}  
\end{align*}
and $\epsilon = e^{i 2\pi/3}$. 
The transformation of these states is summarized as 
\begin{tabular}{c|cccc}
state & ${\cal K}$ & $C_{3z}$ & $C_{2y}$ & $\sigma_{h}$ \\
\hline 
$|\Psi_{1\pm}\rangle$ & 
$\pm|\Psi_{1\mp}\rangle$ 
& $-|\Psi_{1\pm}\rangle$ &
$\pm i|\Psi_{1\pm}\rangle$
& $i|\Psi_{1\pm}\rangle$ \\
$|\Psi_{3+}\rangle$ & 
$|\Psi_{3-}\rangle$ &
$-\epsilon|\Psi_{3+}\rangle$
& $i \epsilon |\Psi_{3-}\rangle $ & $-i|\Psi_{3+}\rangle$\\
$|\Psi_{3-}\rangle$ & 
$-|\Psi_{3+}\rangle$ &
$-\epsilon^2|\Psi_{3-}\rangle$
& $ i\epsilon^2|\Psi_{3+}\rangle$ & $i|\Psi_{3-}\rangle$\\
$|\Psi_{4+}\rangle$ & 
$|\Psi_{4-}\rangle$ &
$-\epsilon^2|\Psi_{4+}\rangle$
& $i\epsilon^2 |\Psi_{4-}\rangle  $ & $-i|\Psi_{4+}\rangle$
\\
$|\Psi_{4-}\rangle$ & 
$-|\Psi_{4+}\rangle$ &
$-\epsilon|\Psi_{4-}\rangle$ &
$i\epsilon|\Psi_{4+}\rangle$ &$i|\Psi_{4-}\rangle$
\\
\end{tabular}
\vspace{.05in}\\
\noindent
where $\sigma_h$ is a reflection through  the $xy$ plane. $|\Psi_1^{(1)}\rangle $ and $|\Psi_1^{(2)}\rangle$ belong to the same irreducible representation and will be mixed by the Hamiltonian.

The general spin-spin Hamiltonian that is invariant under the symmetry group is
\begin{align}
H  =& \sum_{\langle ij\rangle}{\cal J}_1 S_{iz}S_{jz} + {\cal J}_2 S_{ix}S_{jx} \\
&+ {\cal J}_3 S_{iy}S_{jy} 
+ {\cal J}_4[J_{iz}J_{jy} - J_{iy}J_{jz}]
\end{align}
where the sum is over all three pairs of spins.  The $2\times 2$ block for the $|\Psi_{1}^{(1,2)}\rangle$ states is
$$
H_{\Psi_1} = 
\left(\begin{array}{cc}
\frac{3{\cal J}_1}{4} & \frac{\sqrt{3}({\cal J}_2 - {\cal J}_3)}{4} \\
\frac{\sqrt{3}({\cal J}_2 - {\cal J}_3)}{4} & -\frac{{\cal J}_1}{4} + \frac{{\cal J}_2}{2} + \frac{{\cal J}_3}{2}
\end{array}\right)
$$
and the eigenvalues of $|\Psi_3\rangle$ and $|\Psi_4\rangle $  are $E_3 =  -\left(\frac{{\cal J}_1 + {\cal J}_2 + {\cal J}_3}{4}\right) + \frac{\sqrt{3}{\cal J}_4}{2}$
and 
$E_4 = -\left(\frac{{\cal J}_1 + {\cal J}_2 + {\cal J}_3}{4}\right) -\frac{\sqrt{3}{\cal J}_4}{2}$, respectively.  The eigenvectors of the $2\times 2$ block will be of the form
$|\Psi_{1,2}\rangle = \alpha|\Psi_1^{(1,2)}\rangle \pm \beta |\Psi_1^{(2,1)}\rangle$ ($\alpha$, $\beta$ real), and the concurrences of the eigenstates are 
\begin{eqnarray}
    C_1 & = &  \sqrt 2|\beta|\sqrt{1 - \beta^2/3}\\
    C_2 & = &   \sqrt 2|\alpha|\sqrt{1 - \alpha^2/3}\\
    C_3 & = & 2/\sqrt{3}\\
    C_4 & =& 2/\sqrt{3} 
\end{eqnarray}
where $C_i = C(\Psi_i)$.

The general, symmetry-invariant form of the density matrix is 
\begin{align*}
\left(
\begin{array}{rrrrrrrr}
a/2 &  &&  & e&e& e \\ 
& b/6 & c & c & id & -id & & e\\
& c & b/6 & c & -id & & id& e\\
& c &c & b/6 & & id & -id& e\\
e & -id & id &&b/6&c &c & \\
e &id & &-id &c&b/6&c & \\
e & &-id &id &c &c&b/6& \\
 & e&e  & e& & & & a/2 \end{array}\right)\\
 \equiv [a,b,c,d,e].
\end{align*}

The thermal density matrix is
$$
\rho = \sum_i p_i \rho_i 
$$
where 
$$\rho_i = \frac{1}{2}
(|\Psi_{i+}\rangle\langle\Psi_{+}| + |\Psi_{i-}\rangle\langle\Psi_{-}|)
$$
and  $p_i = e^{-E_i/T}/Z$. 
The density matrices of the eigenstates are:
\begin{eqnarray}
\rho_{1}& =& [\alpha^2 ,\beta^2,\beta^2/6,0,\alpha\beta/2\sqrt{3}]\\
\rho_{_2}& =&  [\beta^2,\alpha^2, \alpha^2/6, 0, -\alpha\beta/2\sqrt{3}]\\
\rho_{3} 
&=&[0,1,-1/12, 1/4\sqrt{3},0]\\
\rho_{4}& =& [0,1,-1/12, -1/4\sqrt{3},0]
\end{eqnarray}

Similar to the previous example, we construct a set of symmetric, separable densities, summarized here:\\

\vspace{.05in}
\begin{tabular}{ll}
$\eta_1 =[1,0,0,0,0]$ & $|+++\rangle_{zzz}$ \\
$\eta_2 = [0,1,0,0,0]$ & $|++-\rangle_{zzz}$ \\
$\eta_3 =[1/4,3/4,1/8,0,1/8]$ & $|+++\rangle_{xxx}$ \\
$\eta_4 = [1/4,3/4,1/8,0,-1/8]$ & $|+++\rangle_{yyy}$ \\
$\eta_5 =[1/4,3/4,-1/24,0,1/24]$ & $|++-\rangle_{yyy}$\\
$\eta_6 = [1/4,3/4, -1/24,0,-1/24]$ & $|++-\rangle_{xxx}$\\
$\eta_7 = [1/4,3/4, 0,1/24,0]$ & $|+++\rangle_{xyz}$ \\
$\eta_8 = [1/4,3/4,0,-1/24,0]$ &$|+-+\rangle_{xyz}$ 
\end{tabular} \\
\vspace{.05in}

\noindent
where the last column lists example separable states from which the separable densities can be generated (as always, the lower subscripts refer to local axes). Separable densities that can be expressed as a positive sum of other separable densities are redundant and have been omitted.

As in the previous section, we compute the  optimal decomposition of the thermal density $\rho$,
$$
\rho = \sum_{i=1}^4 p'_i\rho_i + \sum_{j=1}^8 q_i \eta_j
$$
in order to find the concurrence given by
$$
C(\rho) = \sum_i p_i'C_i.
$$
Our results are presented in Section 5.2.

\section{Computational algorithm}
For a given set of model parameters, we begin by selecting one of the separable density arrays $\eta_j$ that is independent of the set of thermal density arrays $\rho_i$. For simplicity, we denote the one we have selected as $\eta_1$.  All of the other separable densities can then be expressed as a linear combination of $\eta_1$ and $\rho_{i}$:
$\eta_{j\geq 2} = \sum_i c_{ji}\rho_i + c_j' \eta_1$.  Then the thermal density can be re-expressed as
$$
\rho = \sum_i p_i' \rho_i +  \sum_j q_i \eta_j
$$
where 
$$
p_i' = p_i - \sum_j q_j c_{ij}
$$ and 
$q_1 = -\sum_j q_j c_j'$.
The function
$f(q_2, q_3, \ldots ) = \sum_i C_i p_i'$ is minimized with respect to the variables $q_{j\geq 2}$, subject to the constraints 
$ 0 \leq q_j, p_i' \leq 1$. The minimum of $f$ is the concurrence $C_s(\rho)$.

\section{Results}
\subsection{Three spins on a line}

\begin{figure}[ht]
\centering
\subfloat{
\includegraphics[width=40mm]{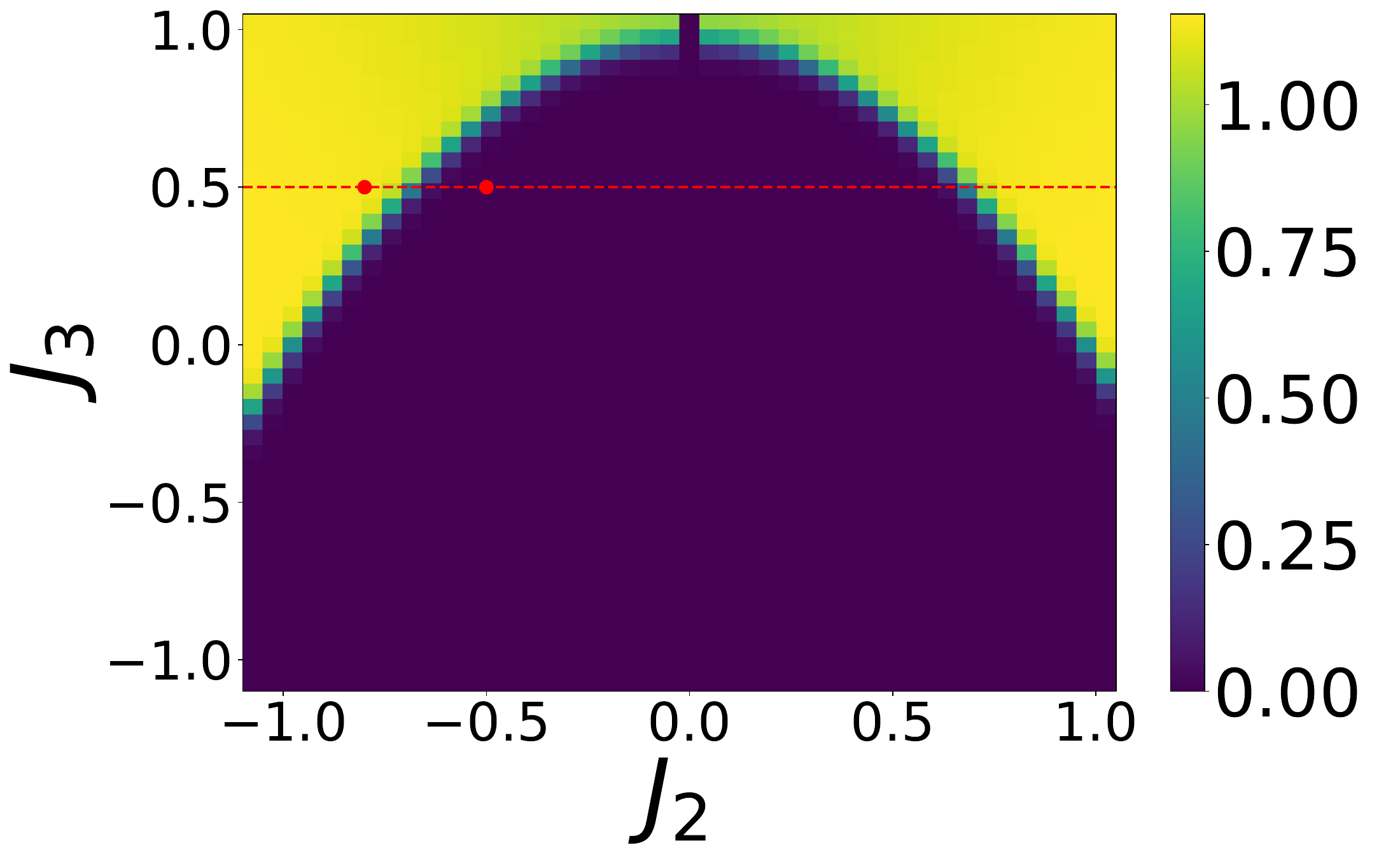}
}
\subfloat{
  \includegraphics[width=40mm]
{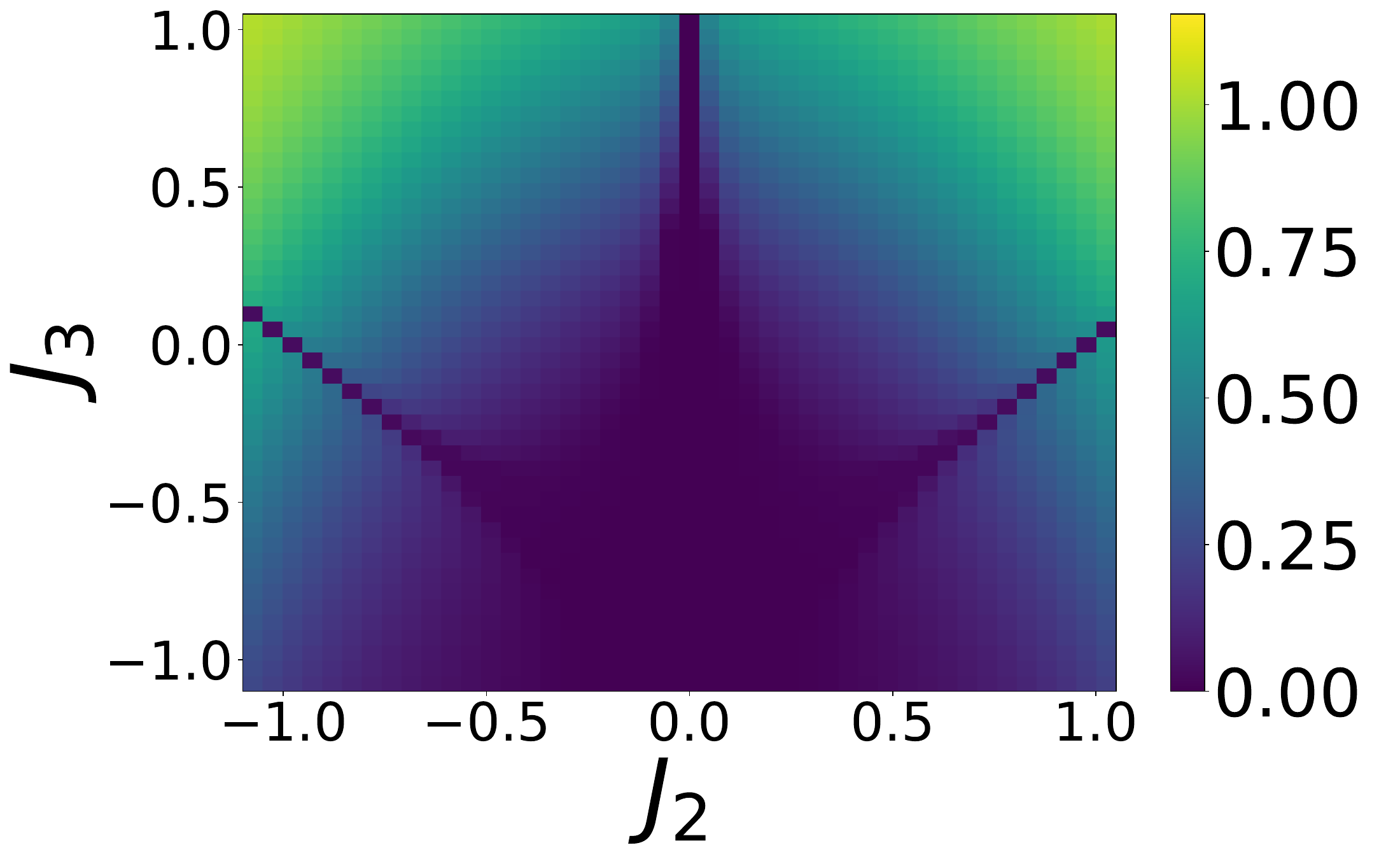}
}
\hspace{0mm}
\subfloat{
  \includegraphics[width=40mm]
{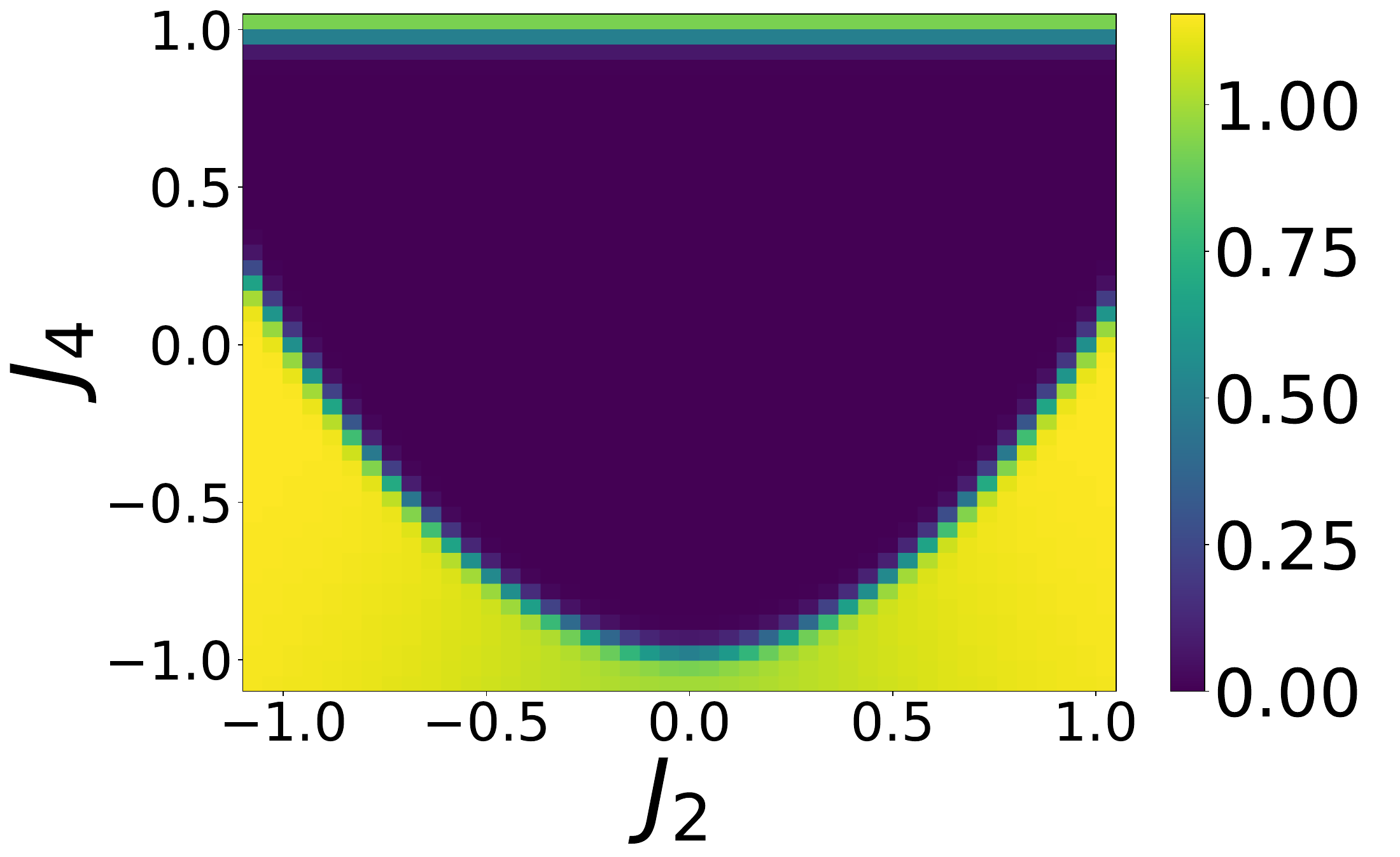}  
}
\subfloat{
  \includegraphics[width=40mm]
{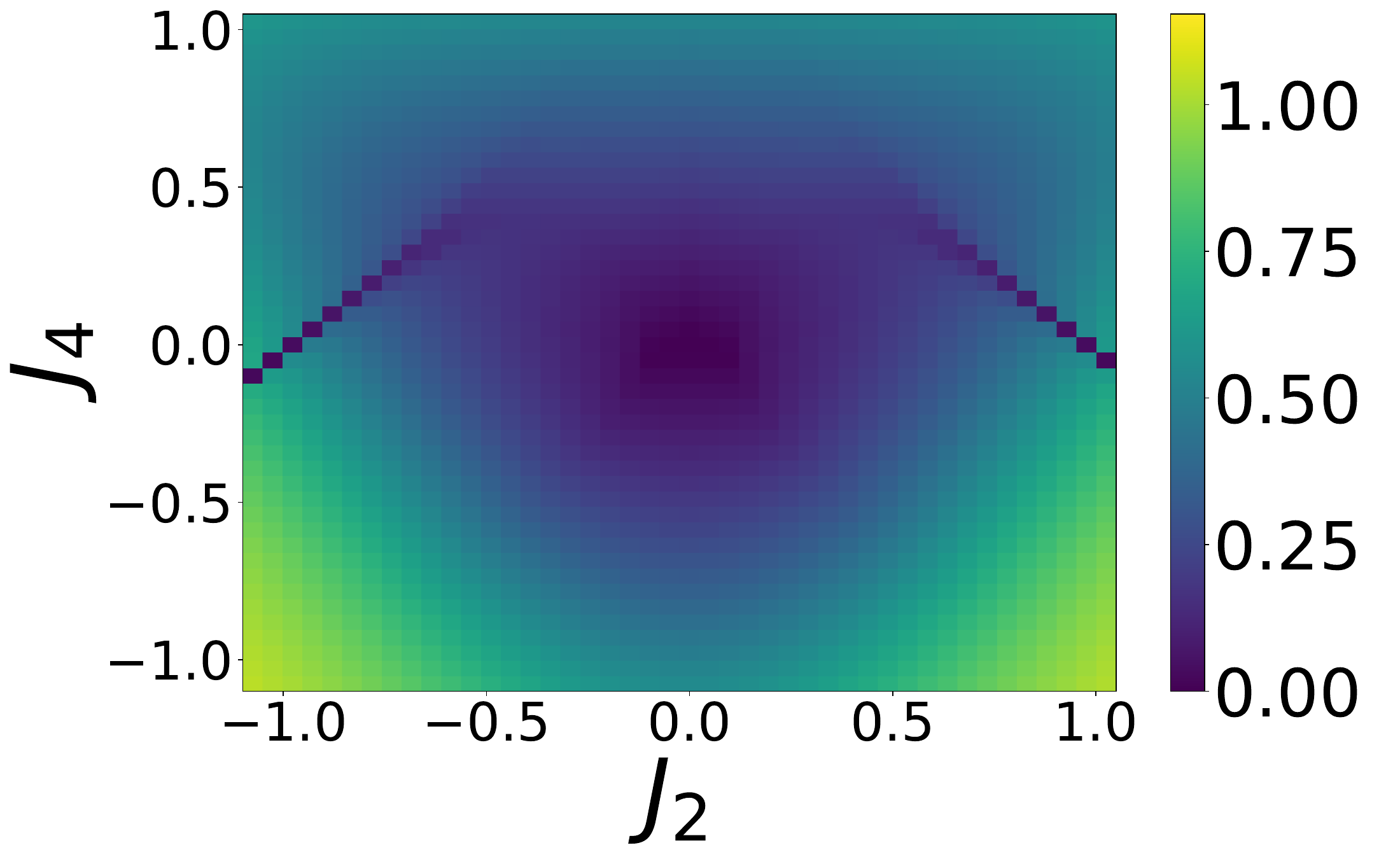}
}
\hspace{0mm}
\subfloat{
  \includegraphics[width=40mm]{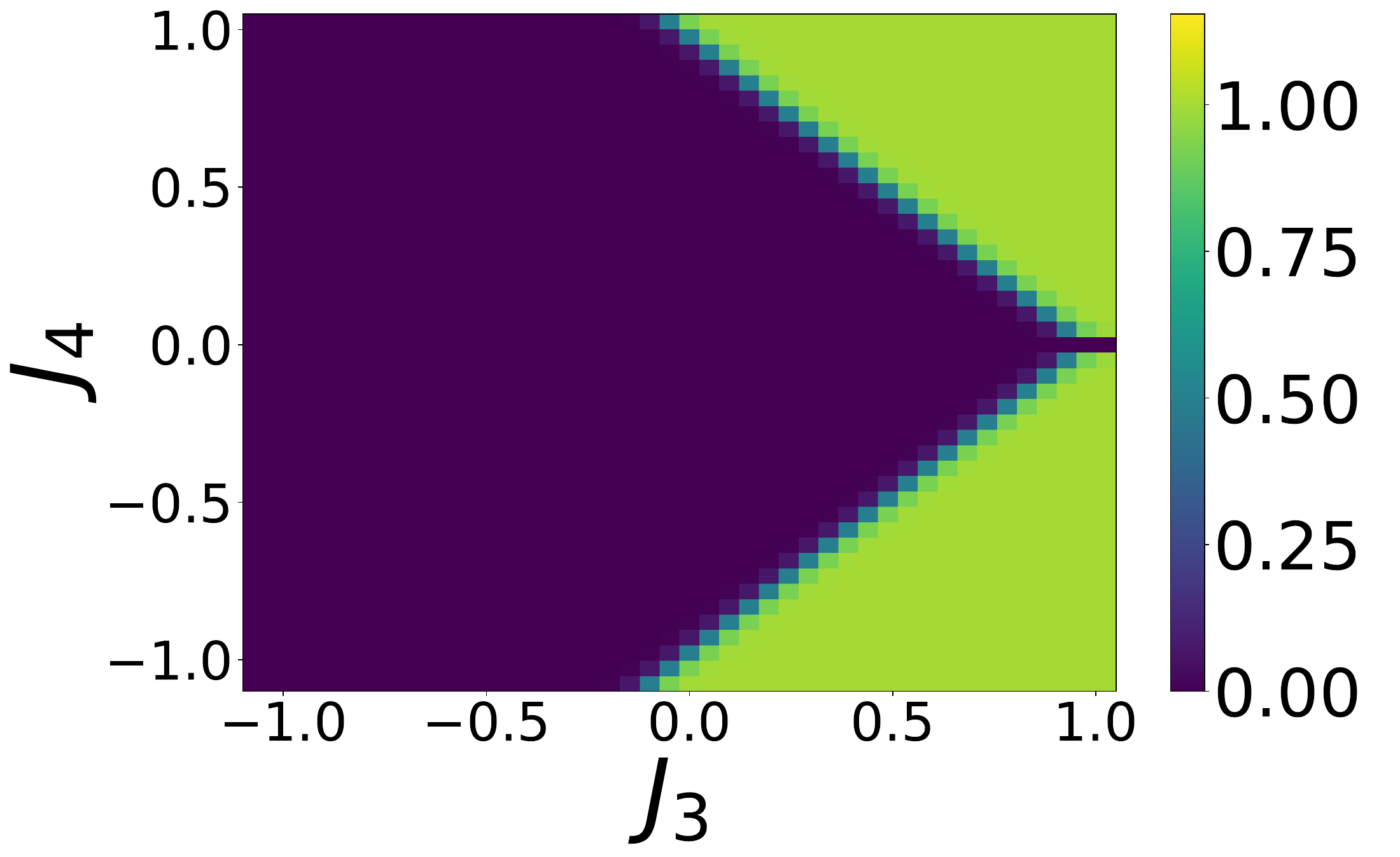}
}
\subfloat{
  \includegraphics[width=40mm]
  {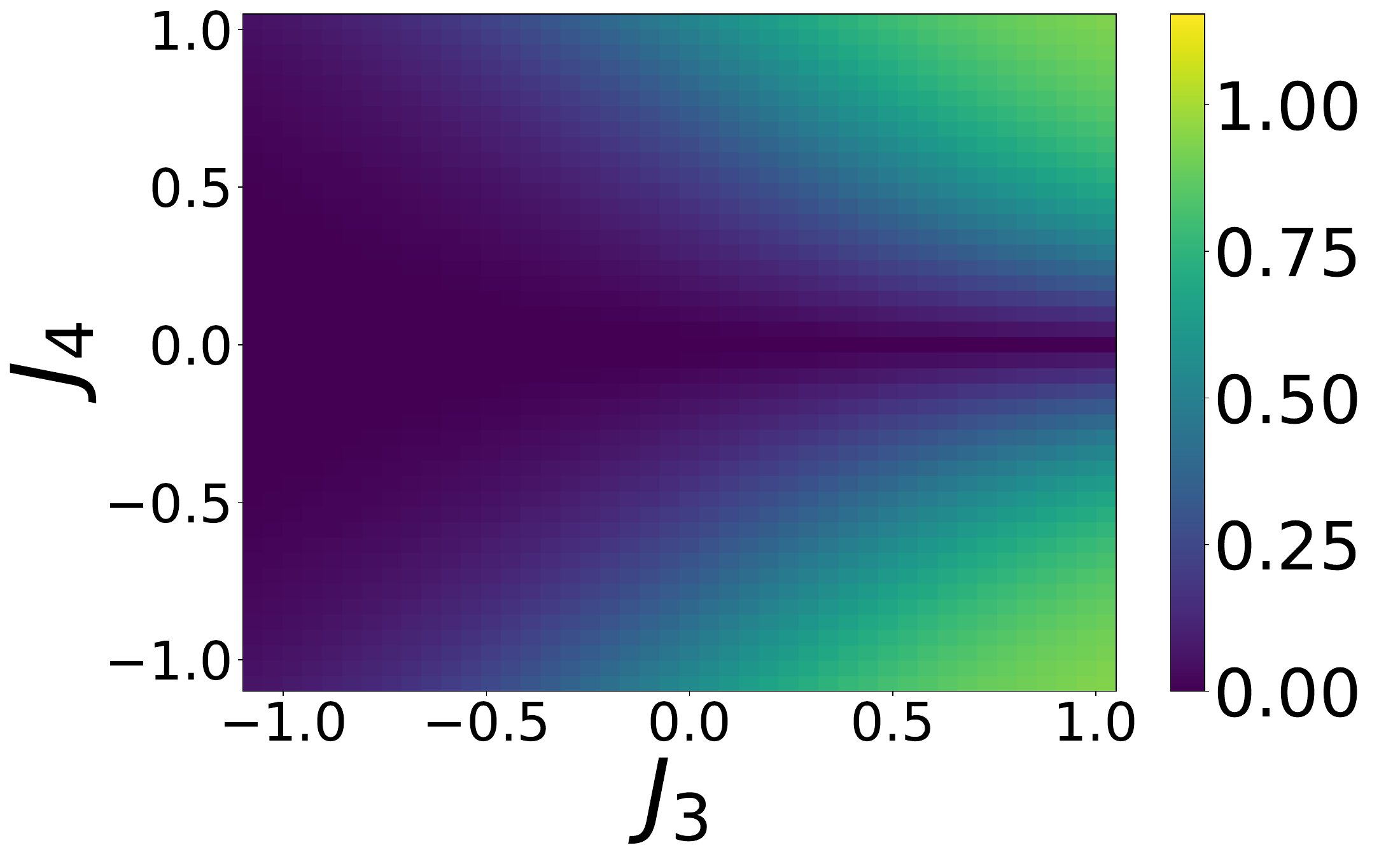}
}
\caption{Concurrence $C_s$ of three spins on a line with ${\cal J}_1 = -1$.  The figures  on the right are for $T=0.01$ and on the left for $T=0.2$. The first row has ${\cal J}_4 = 0$, the second row has ${\cal J}_3 = 0$ and the last row has ${\cal J}_2=0$.
\label{fi:line-FM}}
\end{figure}

The concurrence $C_s$ for a system of three spins on a line is shown in Figs.\ \ref{fi:line-FM} and \ref{fi:line-AF}.
Each figure shows a set of heat maps for two different temperatures, $T=0.01$ and $T=0.2$.  The heat maps plot the concurrence as a function of two out of four of the coupling constants, with ${\cal J}_1=-1$ in Fig.\ \ref{fi:line-FM} and ${\cal J}_1 = 1$ in Fig.\ \ref{fi:line-AF}.

\begin{figure}[ht]
\centering
\begin{minipage}{0.4\textwidth}
\includegraphics[width=\linewidth]{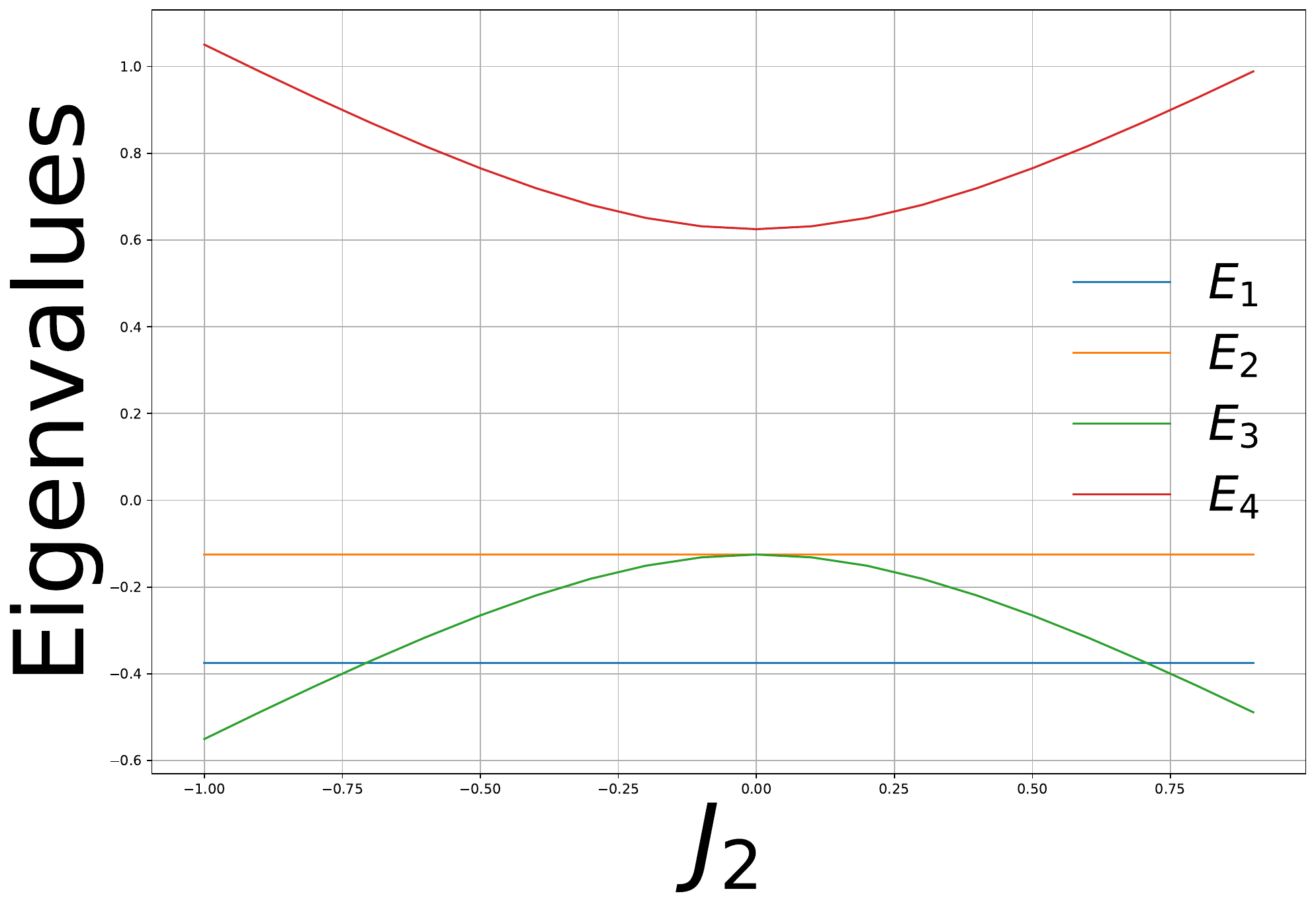}
\end{minipage}
\caption{Eigenvalues of the spin-spin Hamiltonian for three spins on a line as a function of ${\cal J}_2$ for ${\cal J}_1 = -1$, ${\cal J}_3 = 0.5$ and ${\cal J}_4=0$.  A cross-over between 
an unentangled ground state ($E_1$) and an entangled ground state ($E_3$) is evident at 
$J_2 = \pm 0.71$. \label{fi:eigenvals}}
\end{figure}

Let us first consider the centre of each plot shown in Fig.\ \ref{fi:line-FM}, where  the only non-zero coupling constant is ${\cal J}_1 = -1$.  There is a doubly-degenerate ferromagnetic ground state with all spins aligned in the $z$ direction 
 ($|+++\rangle$ and $|---\rangle$), which is obviously separable.  The clear boundaries between vanishing and finite concurrence evident  in the low temperature plots are due to level crossings. Fig.\ \ref{fi:eigenvals} shows the four doubly-degenerate eigenvalues of $H$ along path shown in red in Fig.\ \ref{fi:line-FM}.   Where the concurrence is non-zero it equals the concurrence of the ground state given by Eq. \ref{eq:C4} ; it varies slightly with the coupling constant ${\cal J}_2$ because the parameters $\alpha$ and $\beta$ depend on ${\cal J}_2$.

 In the higher temperature results (right column of Fig.\ \ref{fi:line-FM}), a larger mixture of excited states in the density tends smooth out the boundaries due to level-crossings, increasing the concurrence where it is otherwise vanishing, and decreasing the concurrence elsewhere. 
 The most interesting feature is the narrow regions of vanishing concurrence   along the lines ${\cal J}_3 = \pm {\cal J}_2 -1$.
 Here the two lowest energies are $E_1 = -1/2 + {\cal J}_3/4$ and $E_4 = -1/2 - 3 {\cal J}_3/4$ with associated density $\rho\approx (1-p_4)\rho_1 + p_4 \rho_4$, where $\rho_4 = [0, 2/3, 1/3, 1/6, -1/6]$.  This expression can be exactly re-expressed in terms of separable densities as  
 $\rho = (1-\frac{4}{3} p_4)\eta_1 + \frac{4}{3} p_4\eta_9$, a decomposition that is valid only when $ p_4 \leq \frac{3}{4}$ (that is
 for $E_1 - E_4 = {\cal J}_3  \leq T\,{\rm ln} 3$). Thus the lines of zero concurrence appearing in the top right graph in Fig.\ \ref{fi:line-FM} only extend to values of ${\cal J}_3 \approx 0.21$.

\begin{figure}[ht]
\centering
\begin{minipage}{0.4\textwidth}
\includegraphics[width=\linewidth]{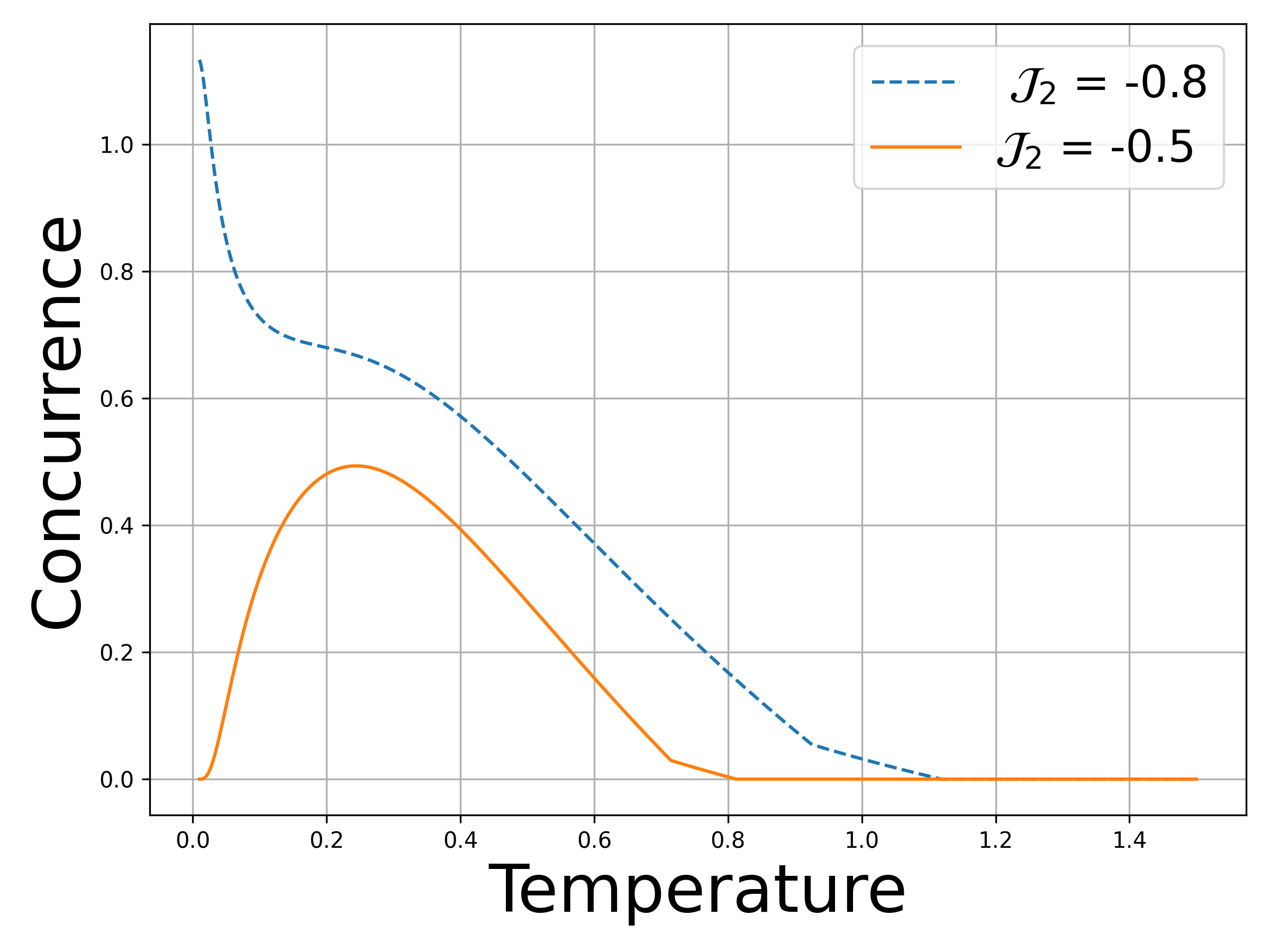}

\end{minipage}
\caption{Concurrence $C_s$ of three spins on a line versus temperature
for parameters ${\cal J}_1 = -1$, ${\cal J}_3 = 0.5$,  ${\cal J}_4=0$, and 
${\cal J}_2 = 0.5$ (lower, solid blue) and ${\cal J}_2 = 0.8$ (upper, dashed red)
.
\label{fi:temp}
}
\end{figure}
 
A plot of concurrence $C_s$ versus temperature is shown in Fig.~\ref{fi:temp} for parameters ${\cal J}_1 = -1$,  ${\cal J}_3 = 0.5$, ${\cal J}_4=0$, and ${\cal J}_2 = -0.5$ and $ -0.8$, indicated by the red dots on the top left plot in Fig.\ \ref{fi:line-FM}.  In the top curve at $T=0$ the concurrence of the thermal density is the concurrence of the ground state, which is non-zero.  The concurrence decreases with temperature, reaching zero (with a discontinuity of the slope) at $T \approx 0.9$. 
The lower curve the ground state has zero concurrence, and so the concurrence of the thermal density vanishes
$T=0$, however it peaks at $T\approx 2$ due to the thermal mixing of entangled states, and then decreases to zero in the large temperature limit.  In this case there are two discontinuities in the slope.  As discussed in Section 3.1.2, such discontinuities occur at a temperature below which terms in the thermal density can be exactly re-expressed in terms of a separable density.

Fig.\ \ref{fi:line-AF} shows a similar set of heatmaps as in Fig.~\ref{fi:line-FM}, except here ${\cal J}_1 = 1$.  At the centre of the plots the four states with two spins up and one spin down or two spins down and one spin up are degenerate, and the concurrence vanishes.  The bottom row of Fig.\ \ref{fi:line-AF} shows sharp transitions in the concurrence that are due to level crossings.

\begin{figure}[ht]
\centering
\subfloat{
\includegraphics[width=40mm]{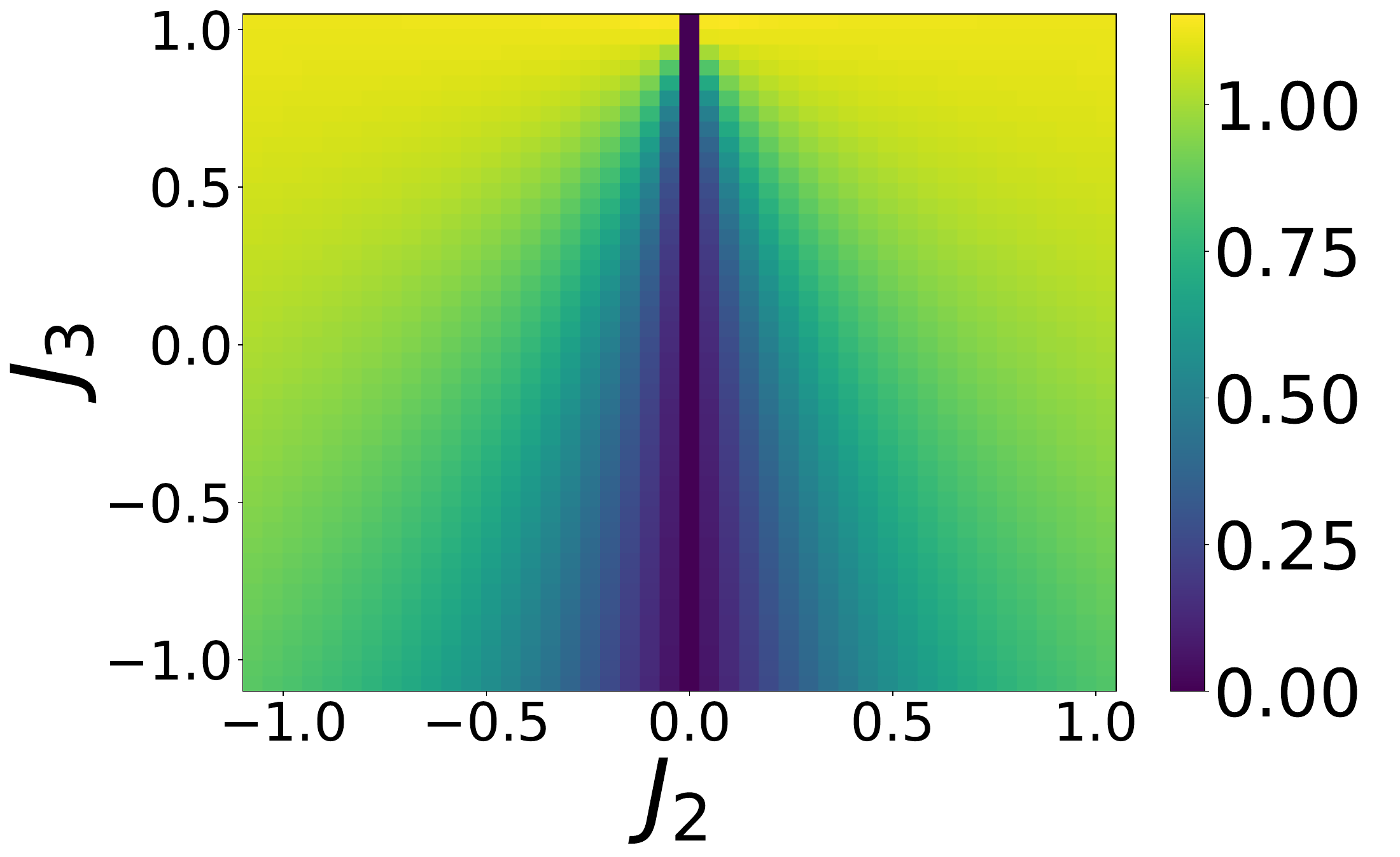}
}
\subfloat{
  \includegraphics[width=40mm]
{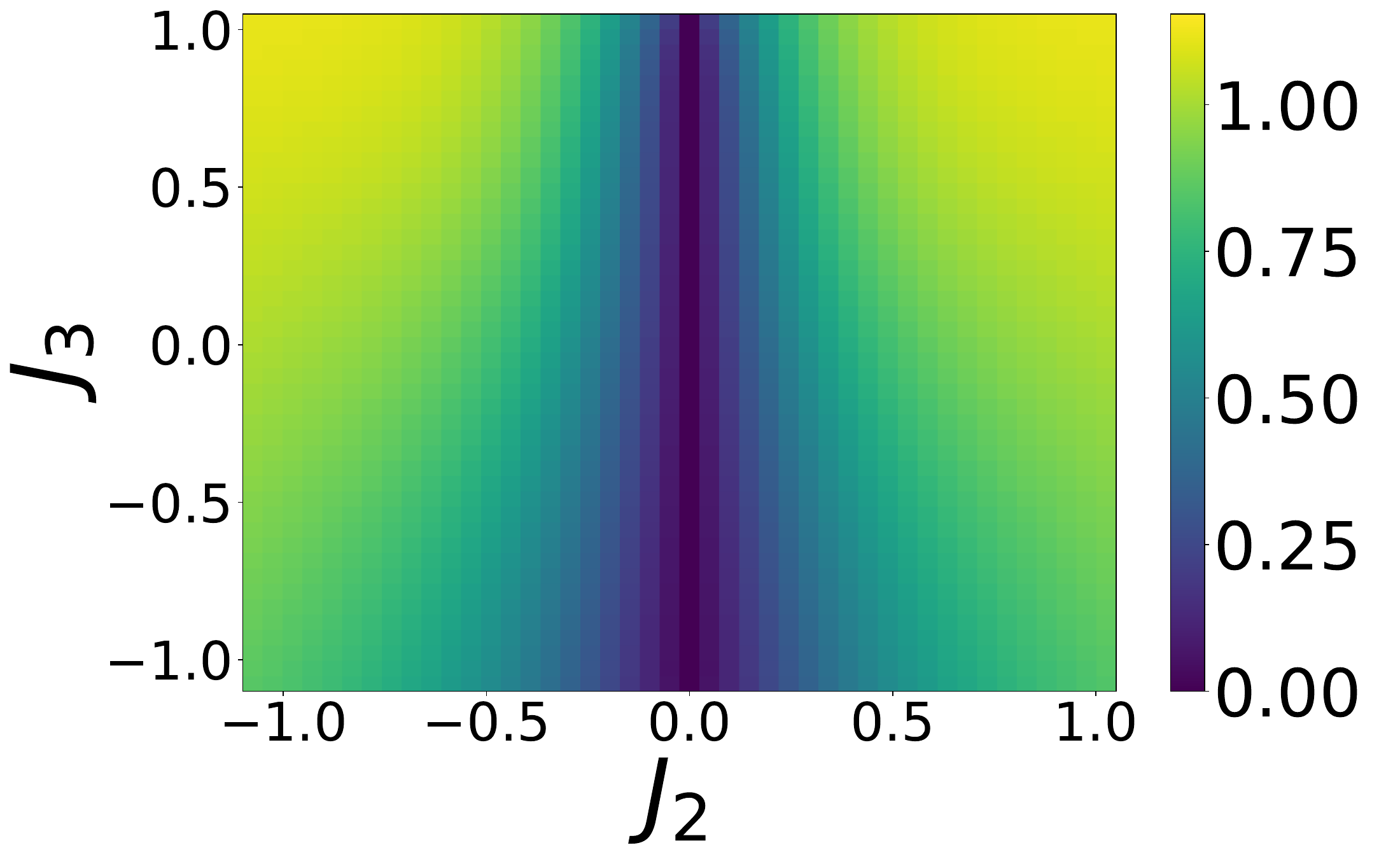}
}
\hspace{0mm}
\subfloat{
  \includegraphics[width=40mm]
{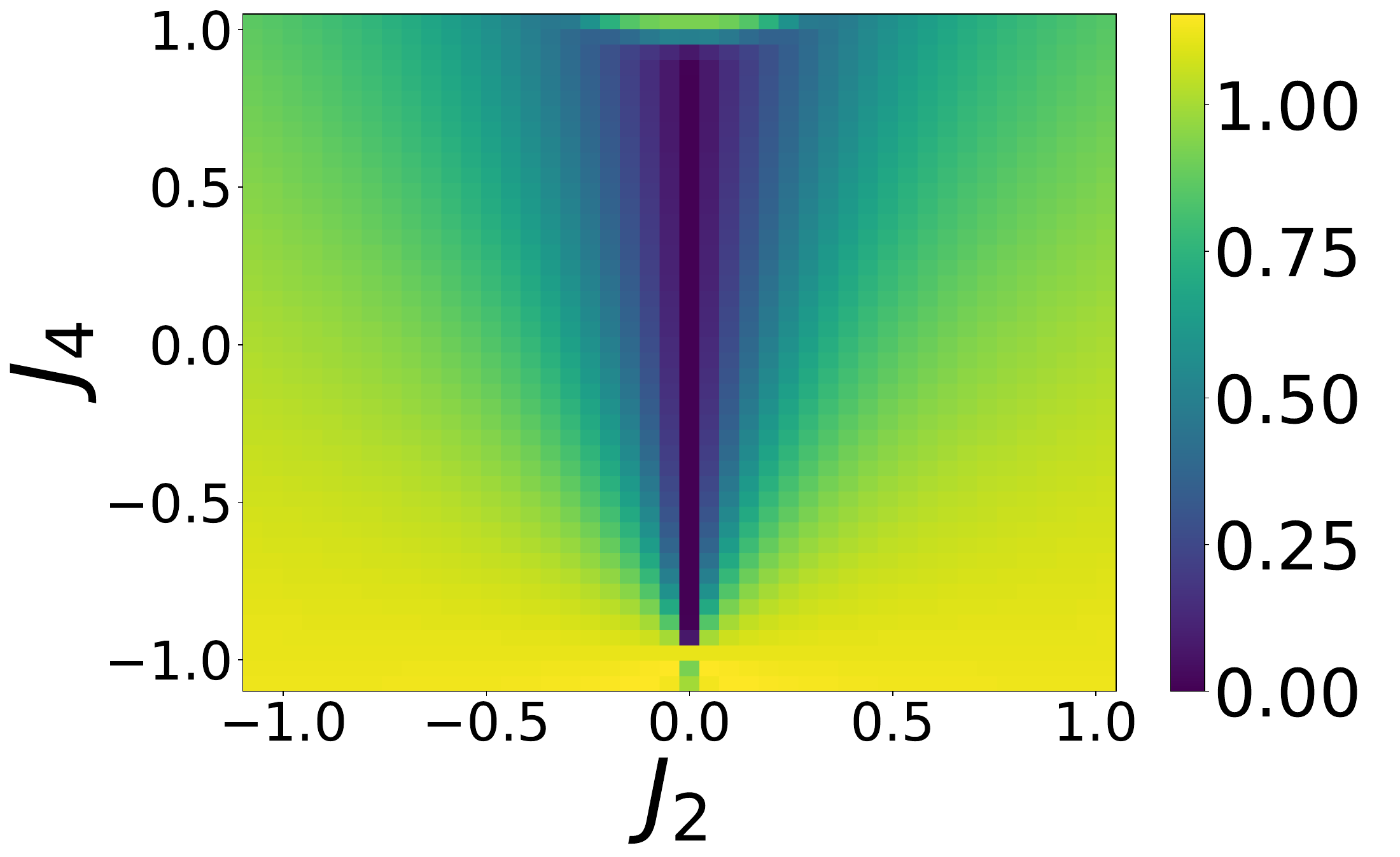}  
}
\subfloat{
  \includegraphics[width=40mm]
{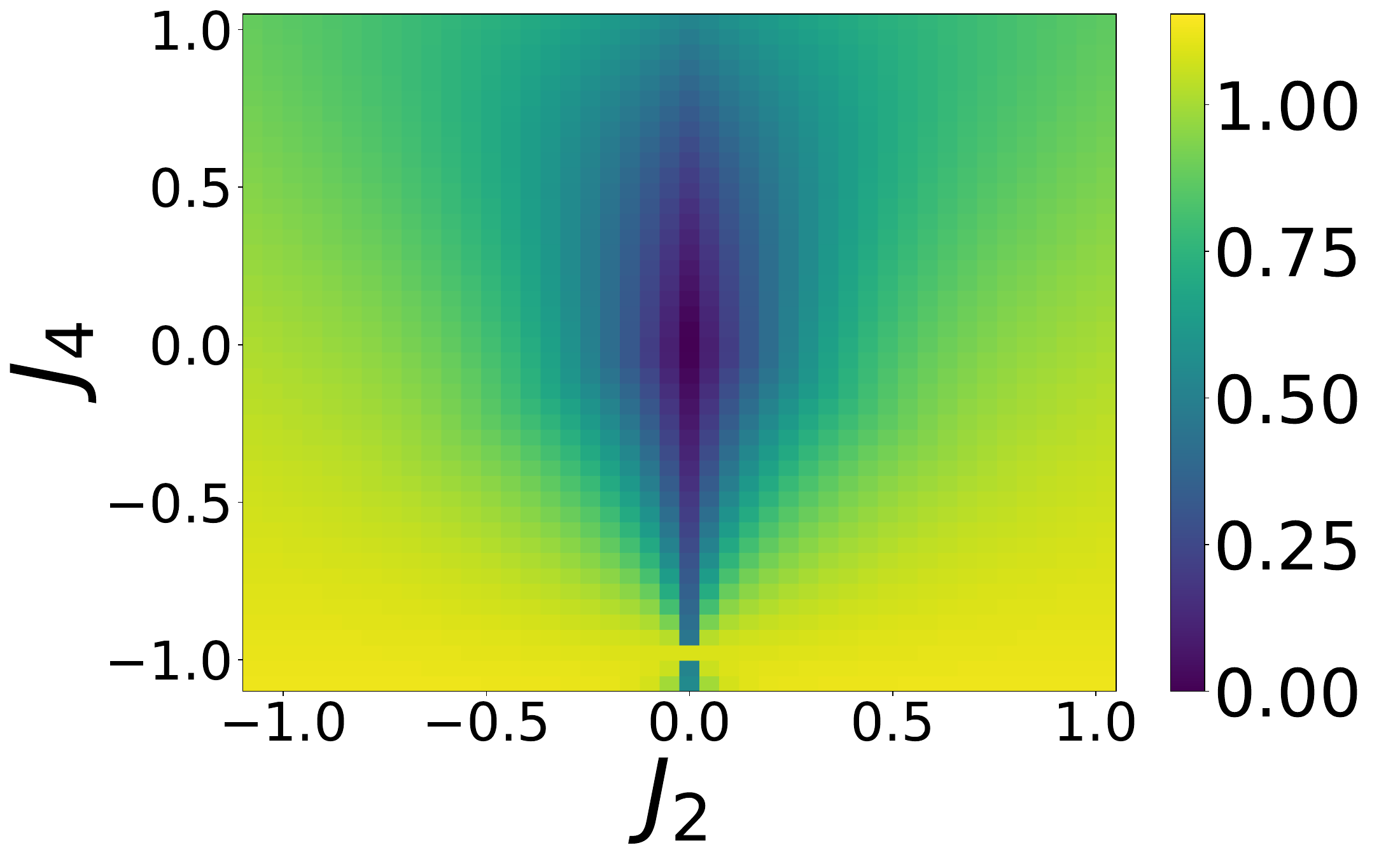}
}
\hspace{0mm}
\subfloat{
  \includegraphics[width=40mm]{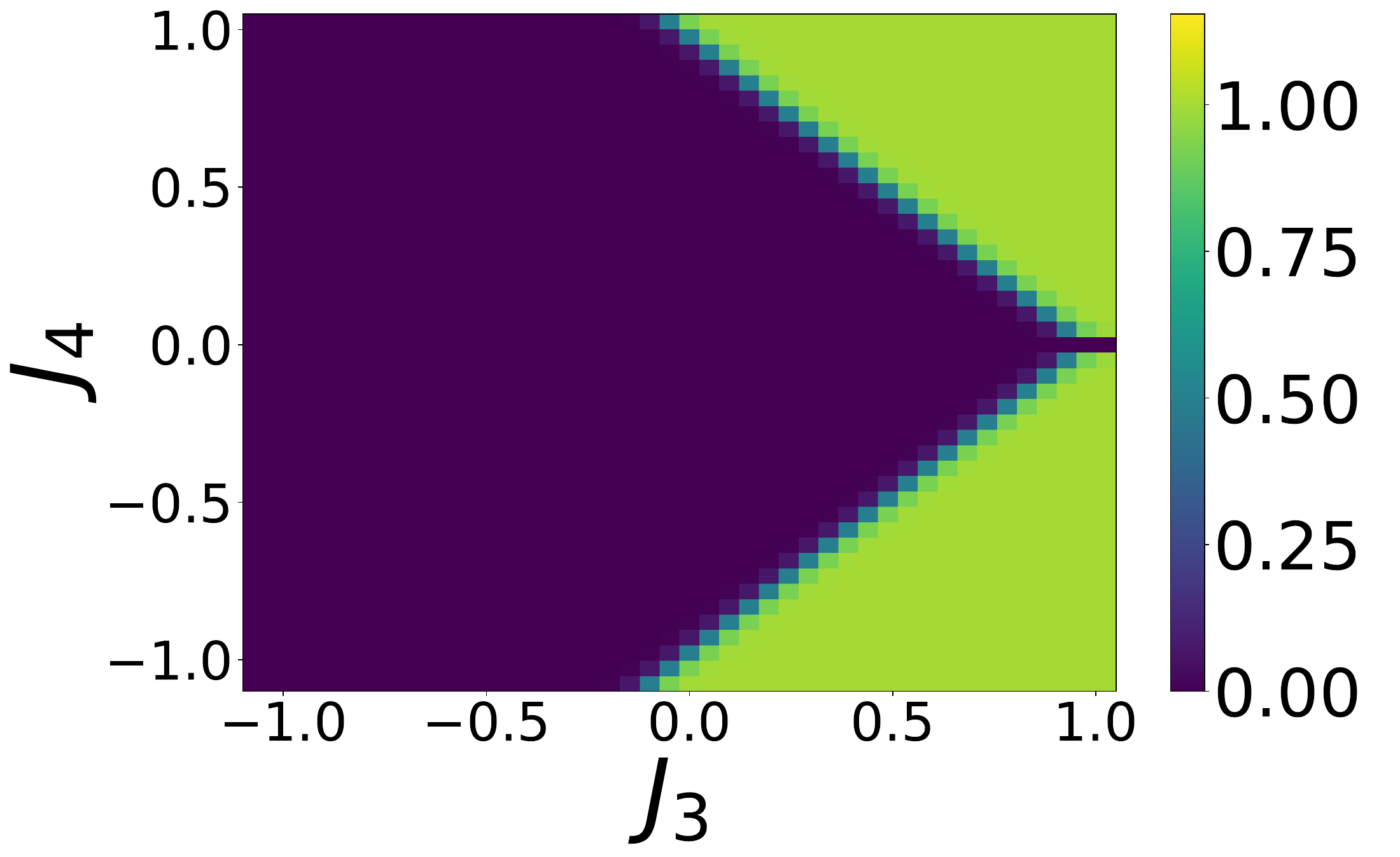}
}
\subfloat{
  \includegraphics[width=40mm]
  {Data2/line_heatmap_j3_j4_T_02.pdf}
}
\caption{Concurrence $C_s$  of three spins on a line for ${\cal J}_1 = 1$.  The figures  on the right are for $T=0.01$ and on the left for $T=0.2$. The first row has ${\cal J}_4 = 0$, the second row has ${\cal J}_3 = 0$ and the last row has ${\cal J}_2=0$.
\label{fi:line-AF}}
\end{figure}

However, the top two rows of Fig.\ \ref{fi:line-AF} show something different. When $|{\cal J}_2|$ becomes non-zero the four-fold degeneracy of the ground state is lifted, the eigenstates will be entangled, and the concurrence becomes non-zero, but unlike the sharp boundaries due to level crossings, here the transition will be smooth, even at zero temperature.  
 This feature can also be seen at the very top, central regions of the plots in the top row of Fig.\ \ref{fi:line-FM} and the rightmost, central regions  of the plots in the bottom row of Figs.\ \ref{fi:line-FM} and \ref{fi:line-AF}.

\subsection{Three spins on a triangle}

\begin{figure}[ht]
\centering
\subfloat{
\includegraphics[width=40mm]{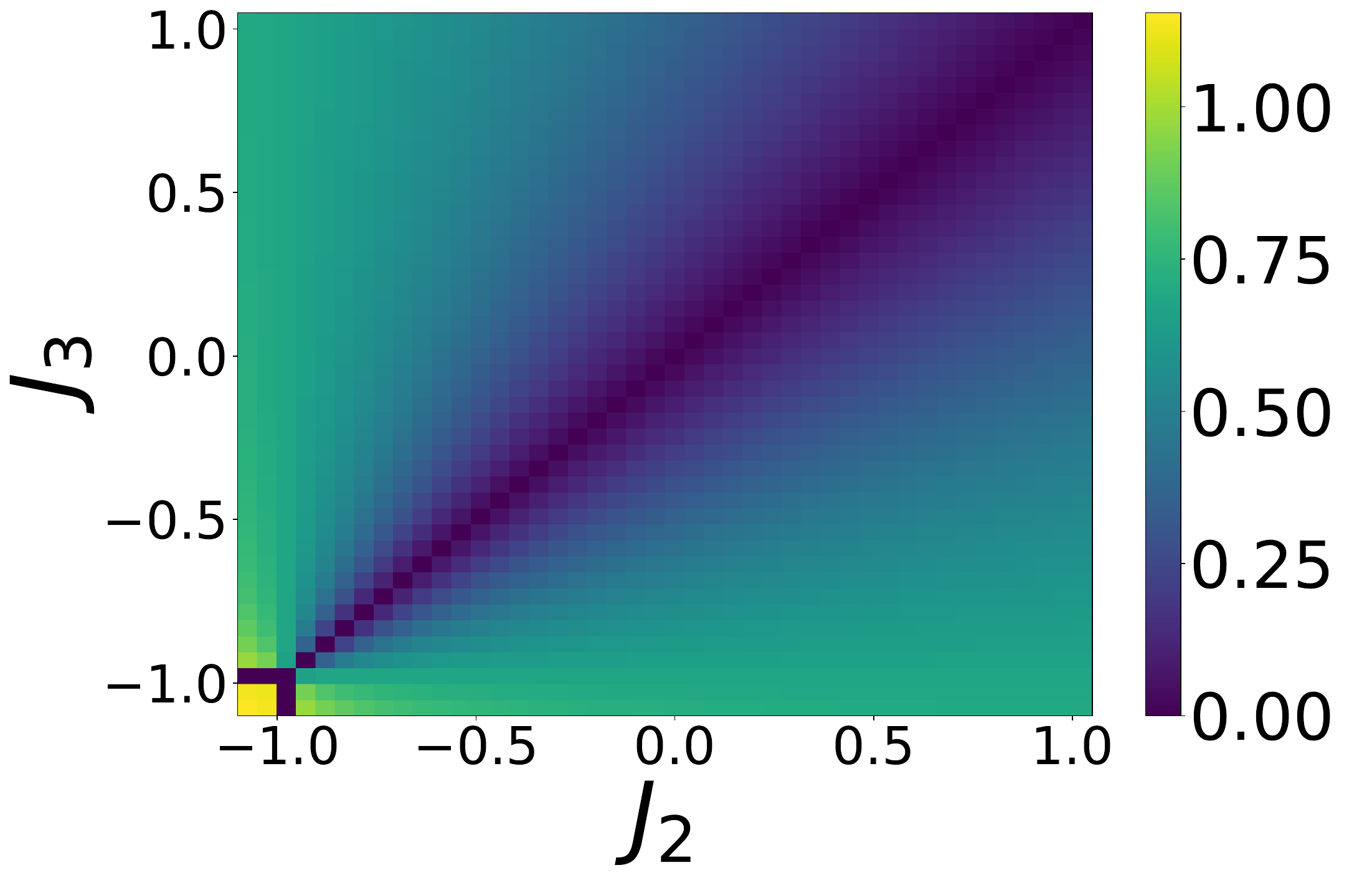}
}
\subfloat{
  \includegraphics[width=40mm]
{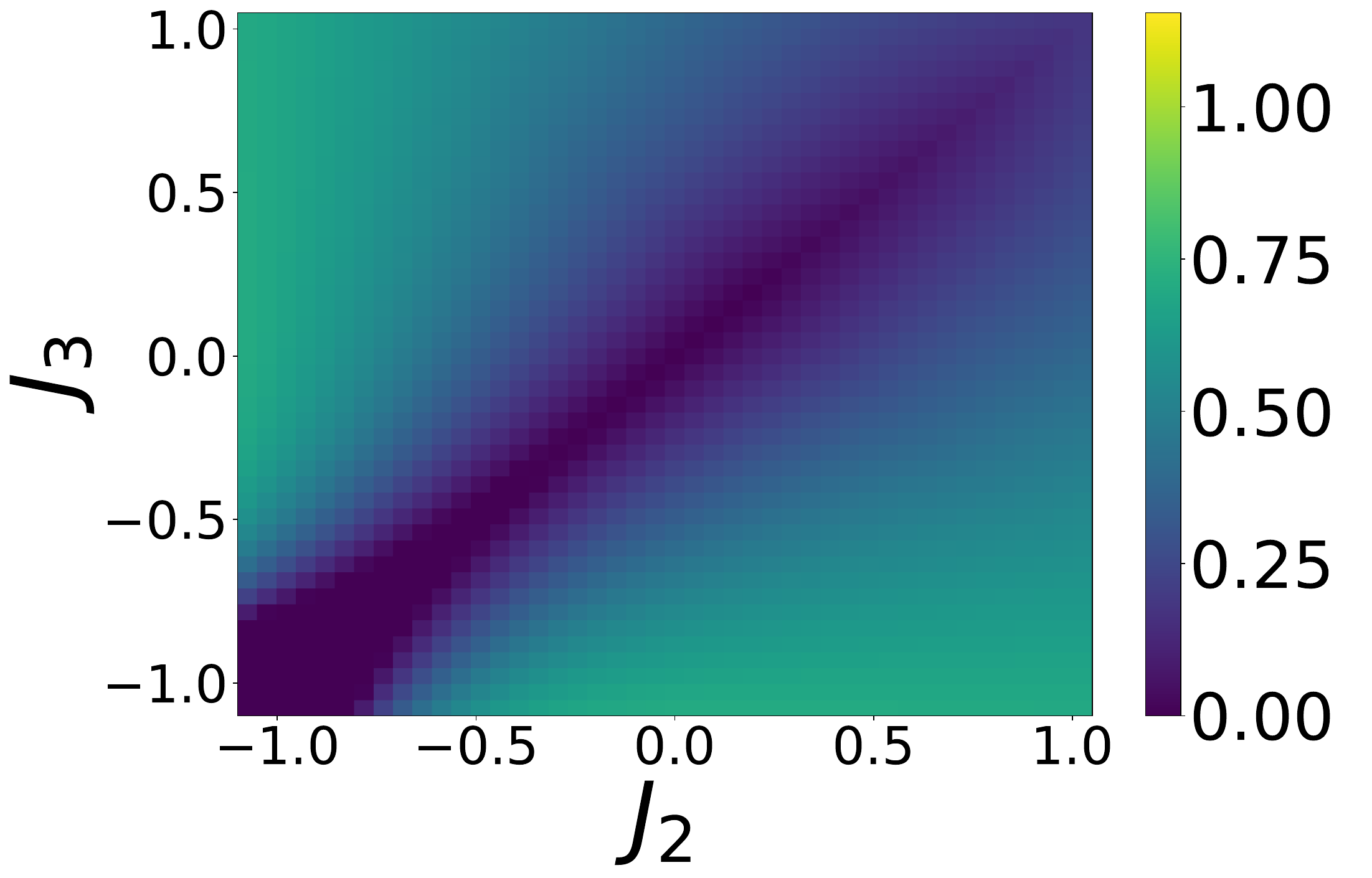}
}
\hspace{0mm}
\subfloat{
  \includegraphics[width=40mm]
{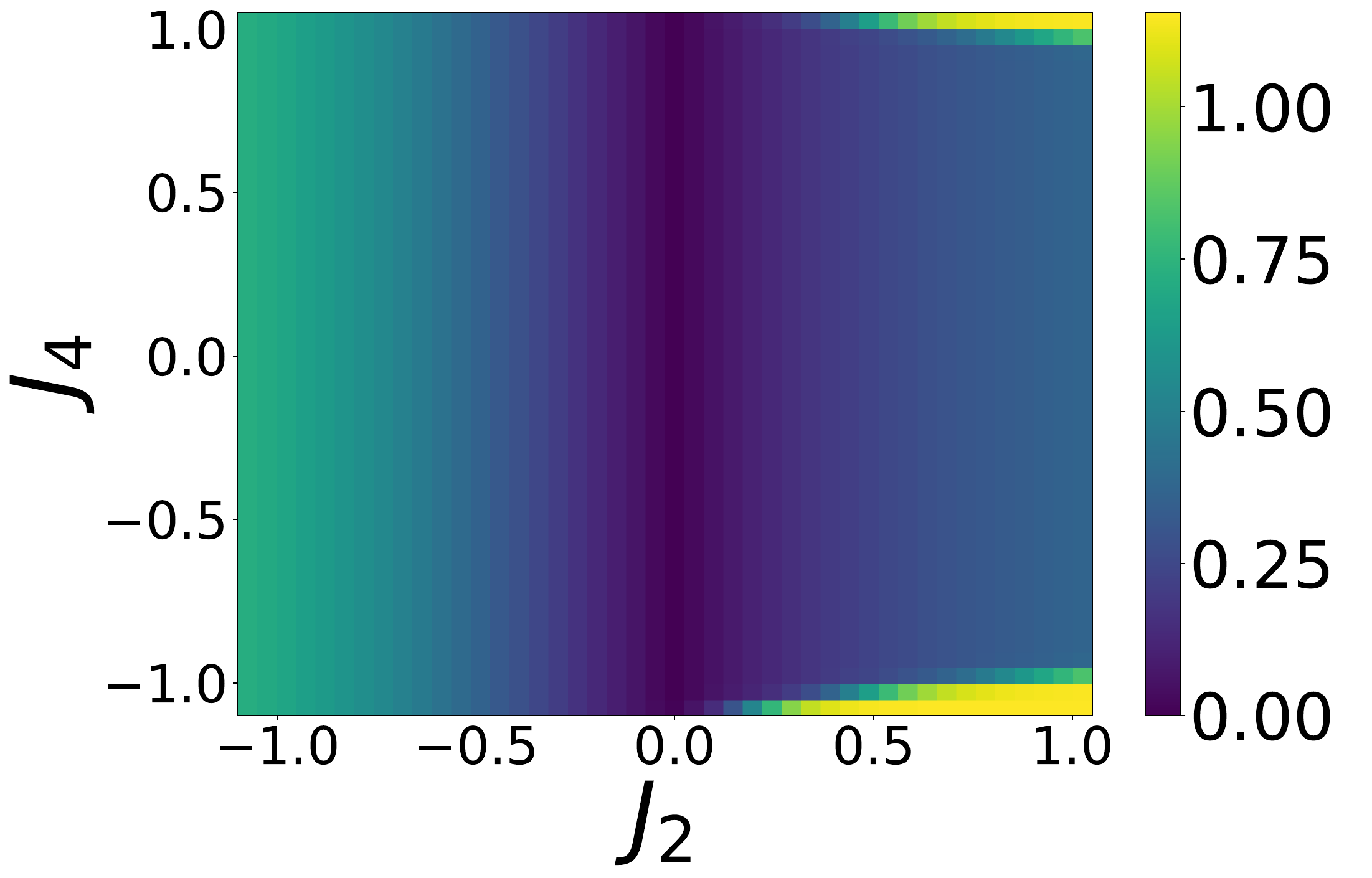}  
}
\subfloat{
  \includegraphics[width=40mm]
{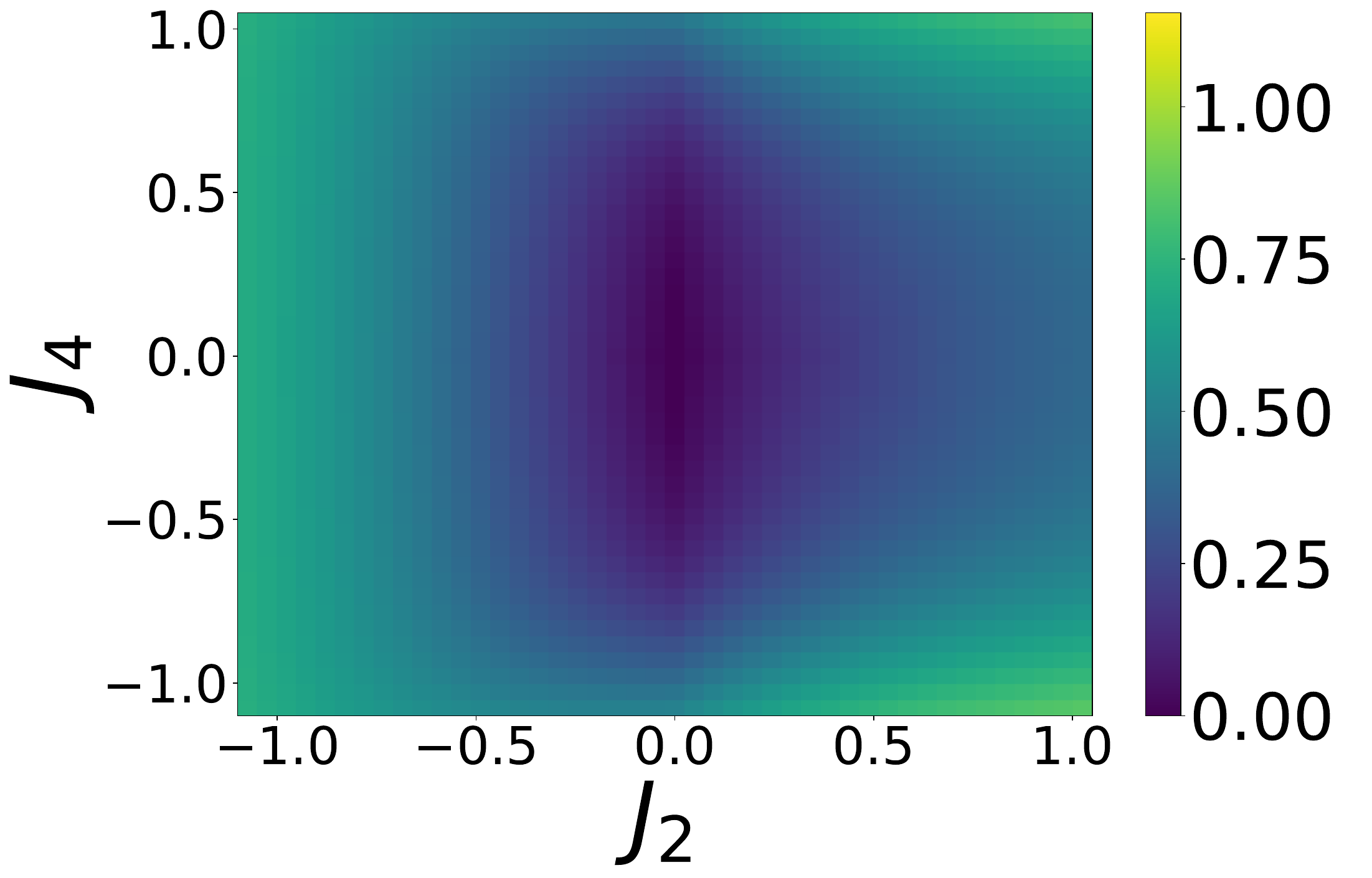}
}
\hspace{0mm}
\subfloat{
  \includegraphics[width=40mm]{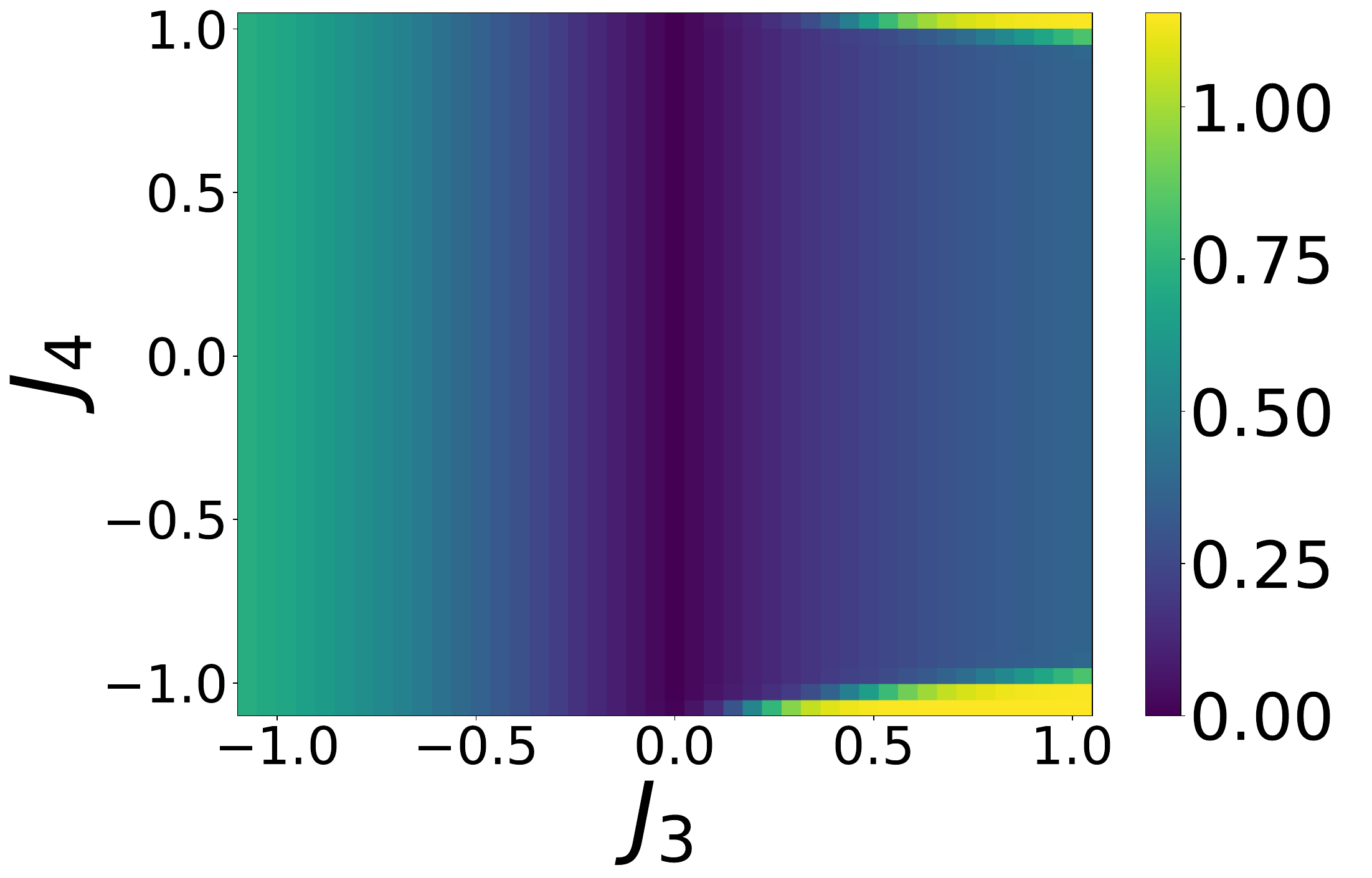}
}
\subfloat{
  \includegraphics[width=40mm]
  {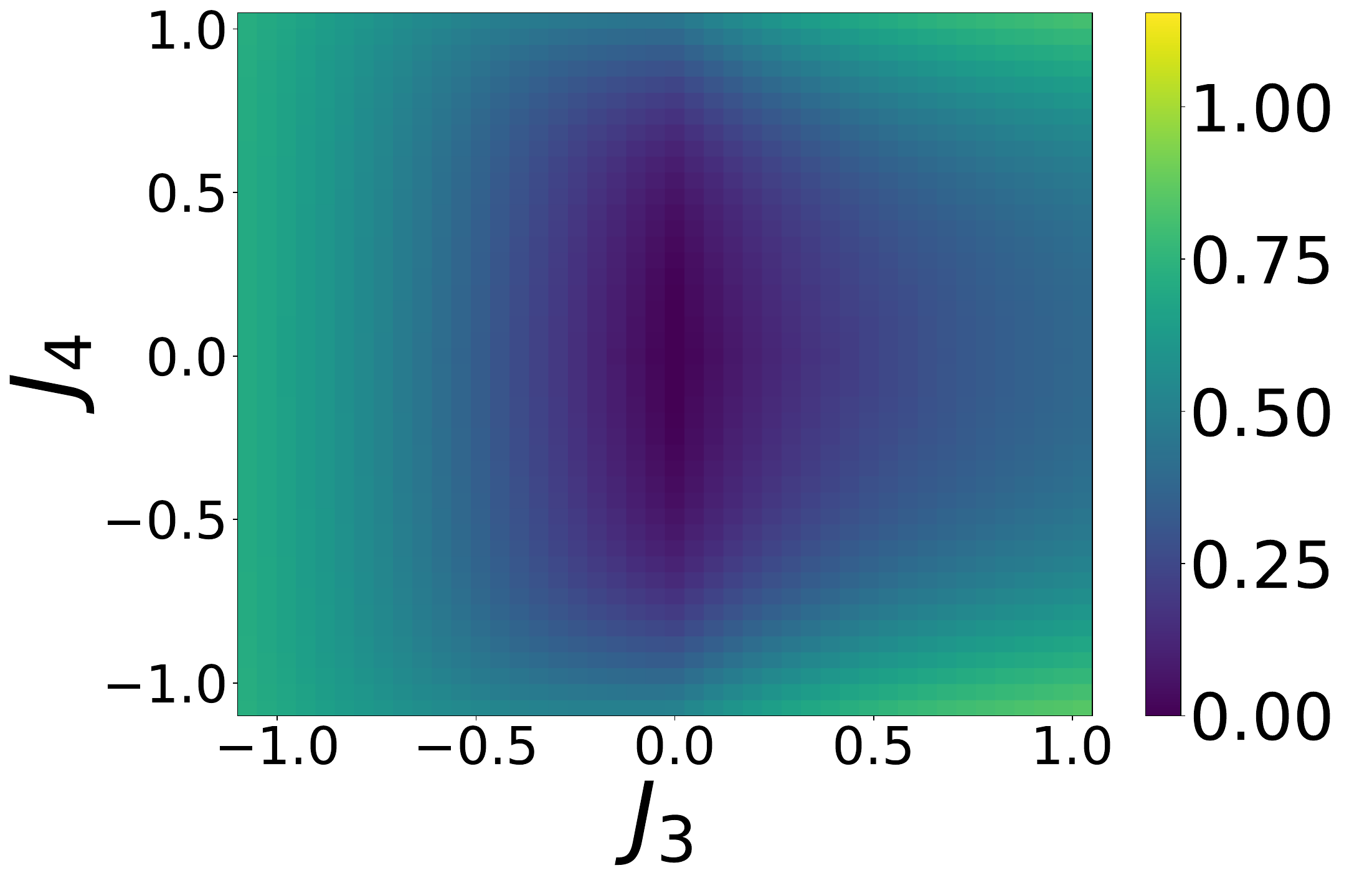}
}
\caption{Concurrence $C_s$ of three spins on a triangle for  ${\cal J}_1 = -1$.  The figures  on the right are for $T=0.01$ and on the left for $T=0.2$. The first row has ${\cal J}_4 = 0$, the second row has ${\cal J}_3 = 0$ and the last row has ${\cal J}_2=0$.
\label{fi:triangle-FM}}
\end{figure}

\begin{figure}[ht]
\centering
\subfloat{
\includegraphics[width=40mm]{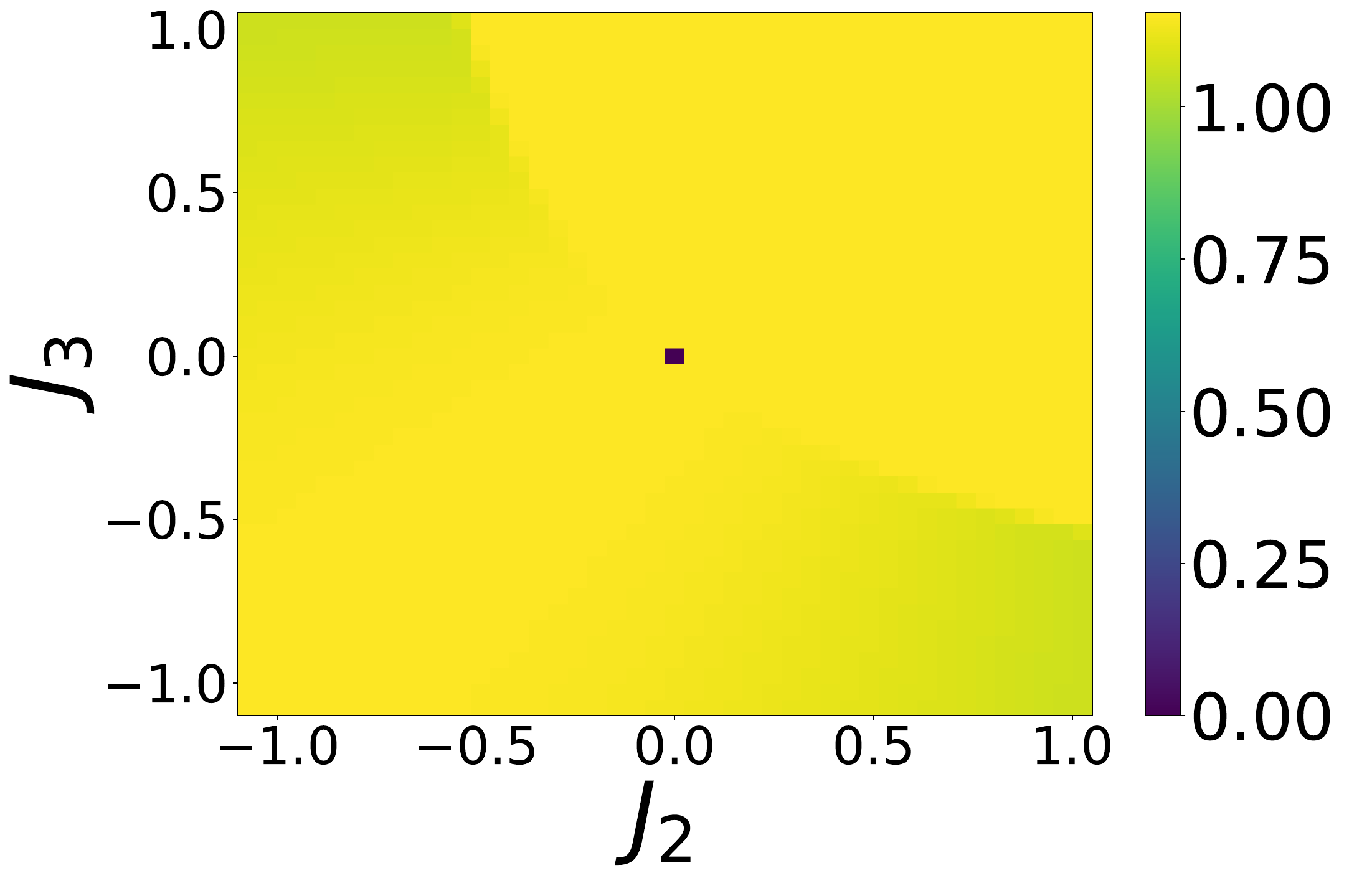}
}
\subfloat{
  \includegraphics[width=40mm]
{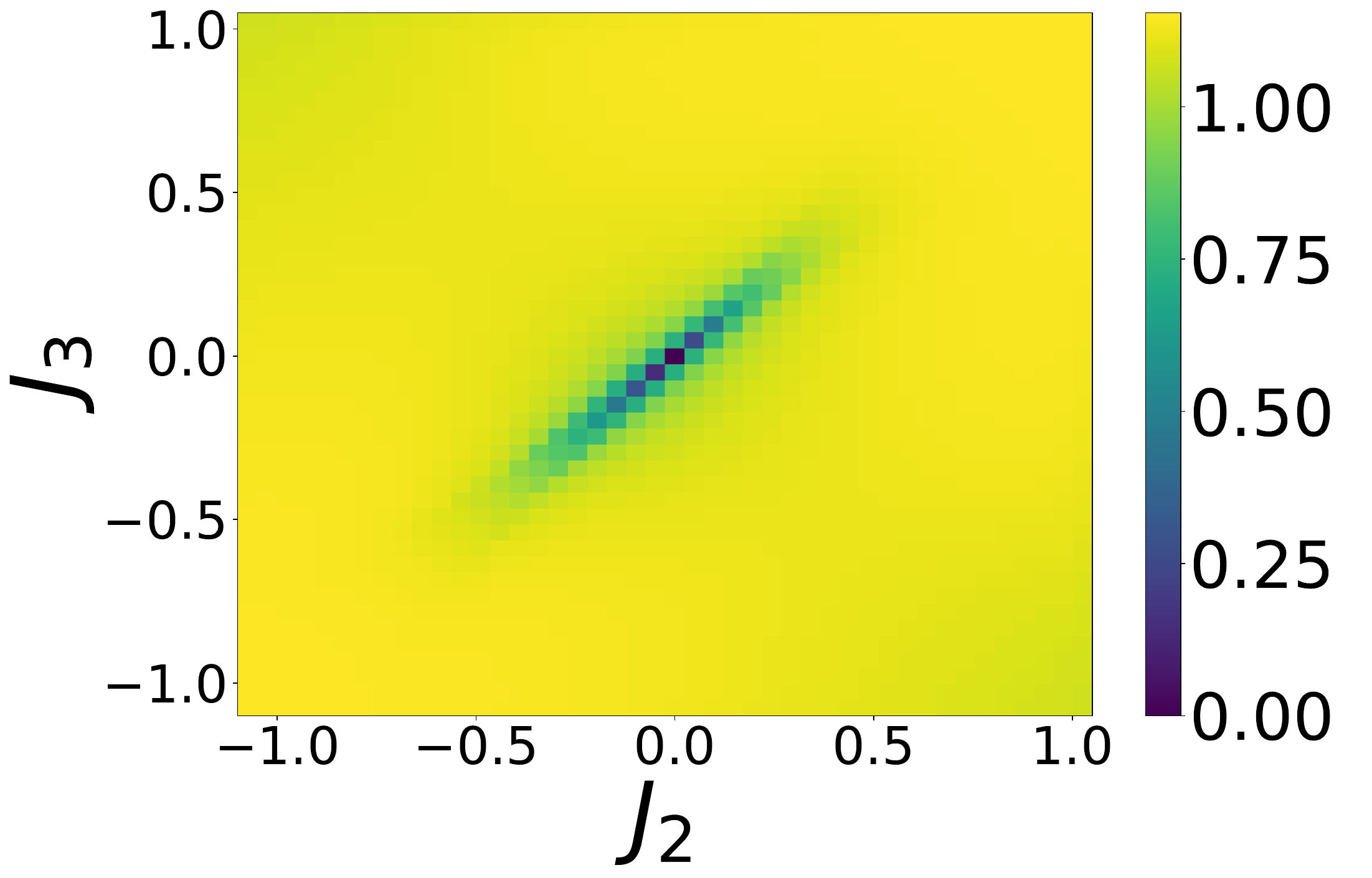}
}
\hspace{0mm}
\subfloat{
  \includegraphics[width=40mm]
{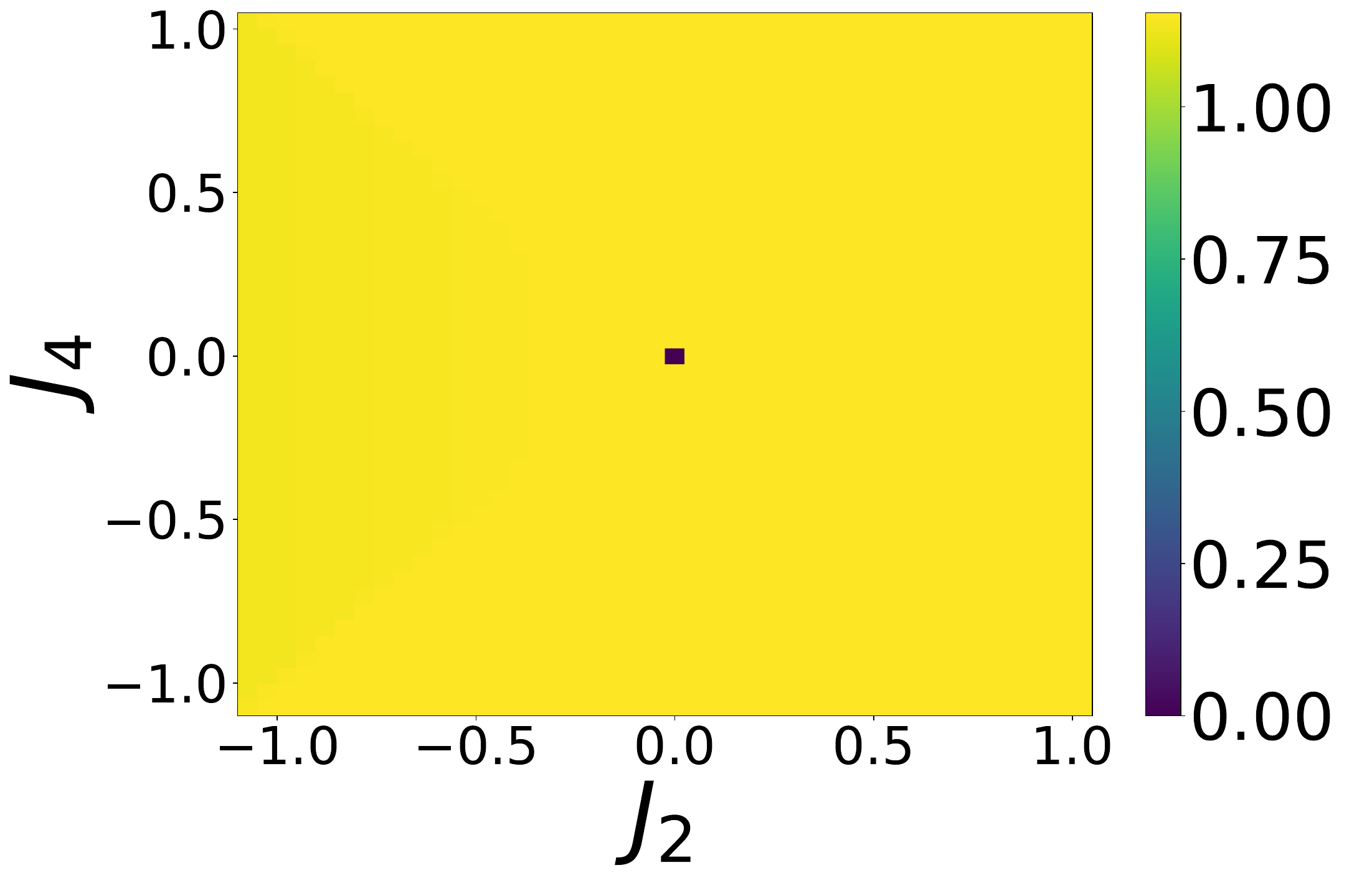}  
}
\subfloat{
  \includegraphics[width=40mm]
{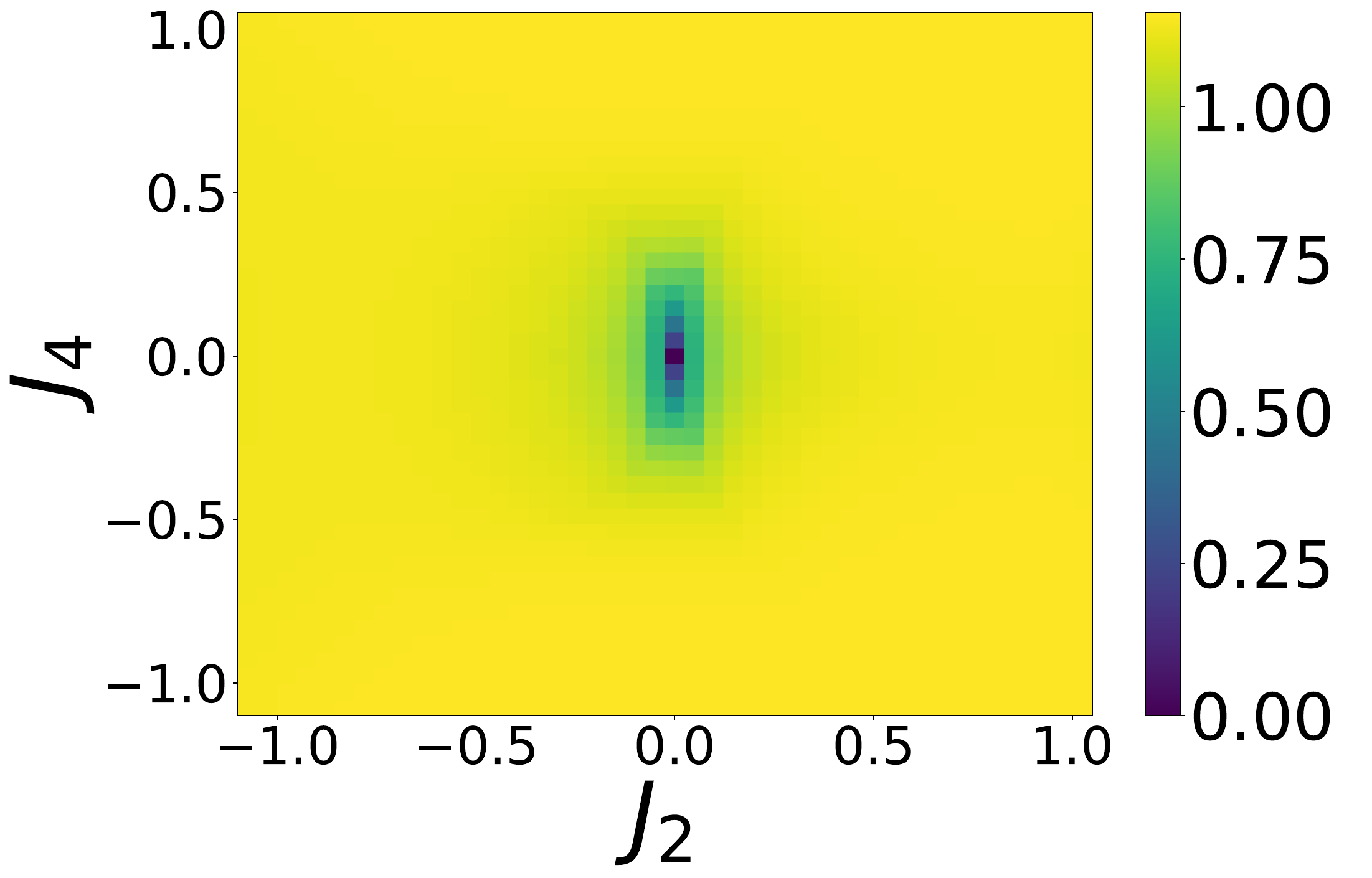}
}
\hspace{0mm}
\subfloat{
  \includegraphics[width=40mm]{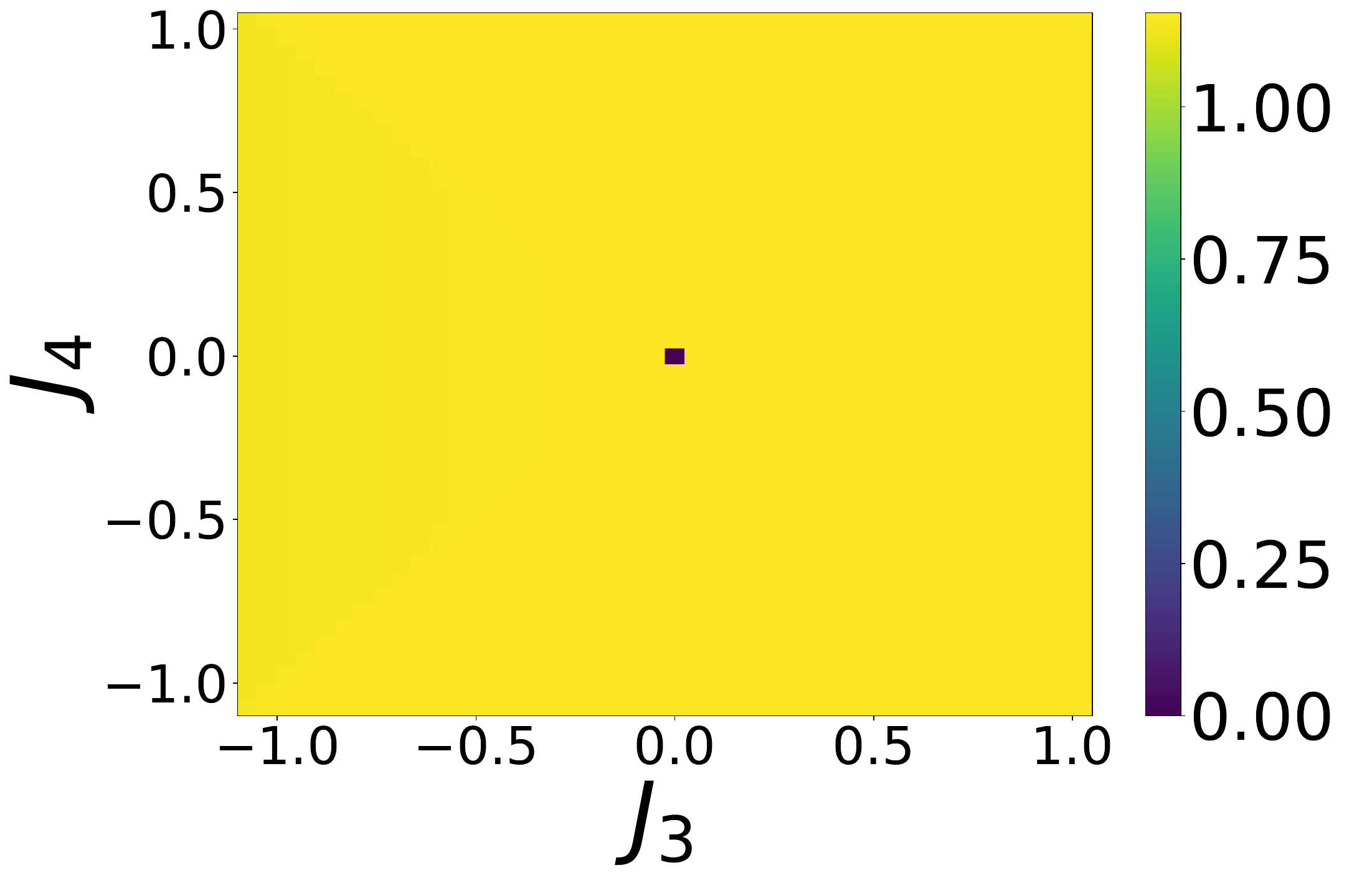}
}
\subfloat{
  \includegraphics[width=40mm]
  {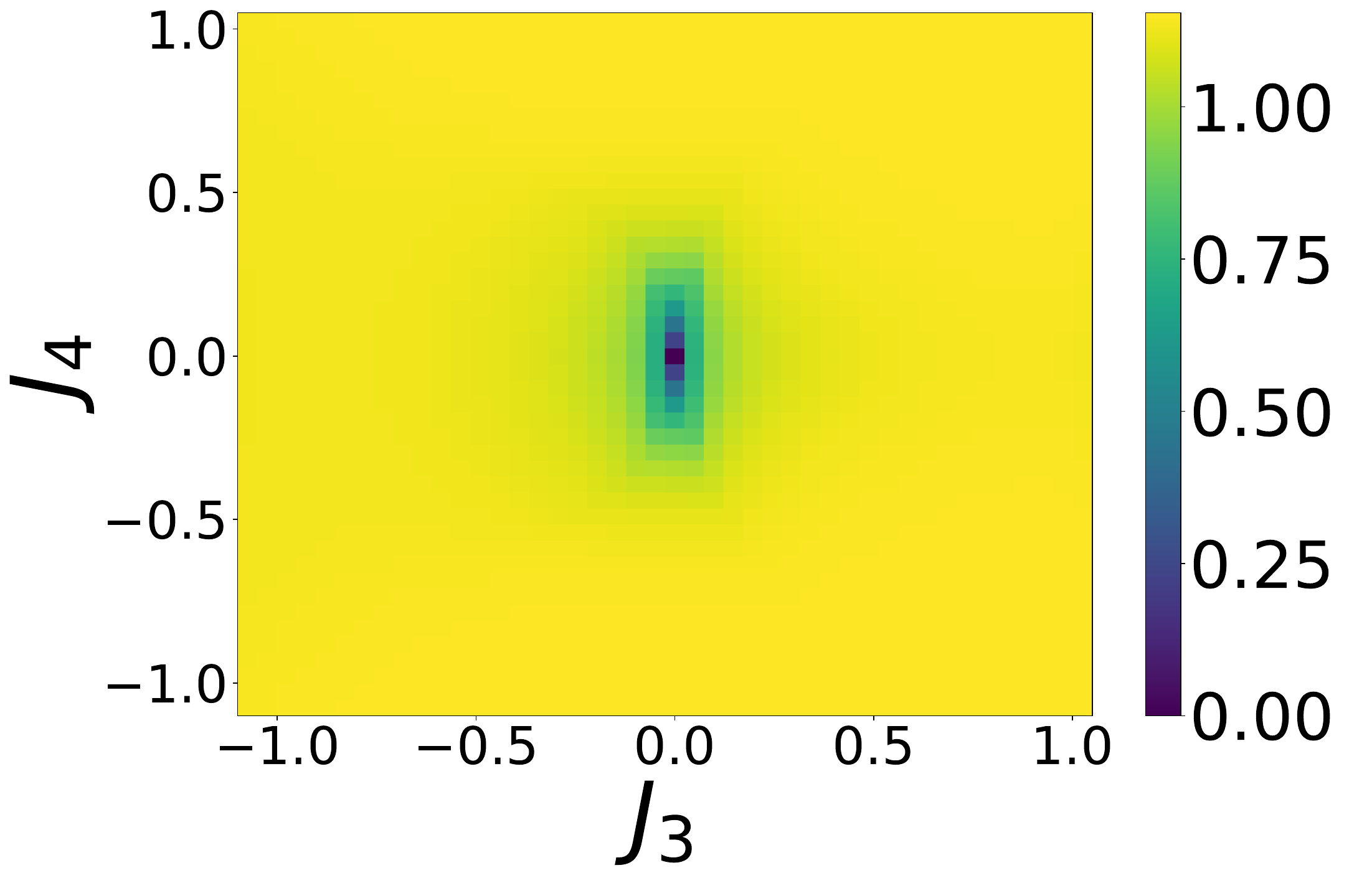}
}
\caption{Concurrence $C_s$ of three spins on a triangle for ${\cal J}_1 = 1$.  The figures  on the right are for $T=0.01$ and on the left for $T=0.2$. The first row has ${\cal J}_4 = 0$, the second row has ${\cal J}_3 = 0$ and the last row has ${\cal J}_2=0$.
\label{fi:triangle-AF}}
\end{figure}

The parameter ${\cal J}_1$ (with all of the others small) controls whether or not the ground state favours the doubly degenerate `all-in-all-out' configuration with all spins pointing into or out of the triangle (${\cal J}_1 <0$) versus the four-fold degenerate `2-in-1-out/1-in-2-out' configurations (${\cal J}_1>0$).   The concurrences for these two cases are shown in  Figs.\ \ref{fi:triangle-FM} and \ref{fi:triangle-AF}.

In Fig.\ \ref{fi:triangle-FM} the ground state is the unentangled doubly-degenerate all-in-all-out state for most of the parameter space shown.  As in the results for spins on a line, the abrupt boundaries found at the edges of the plots are due to level crossings.  Since the separable all-in-out-out states can mix with one of the entangled 2-in-2-out/1-in-2-out configurations as a function of the parameter ${\cal J}_2 - {\cal J}_3$, there will be variations in the ground state concurrence which result in the variations of the density concurrence. The dark bands of vanishing concurrence in all of the plots are where ${\cal J}_2 \approx {\cal J}_3$; here the ground state is the separable all-in-all-out state.

Fig.\ \ref{fi:triangle-AF} show the concurrence for three spins on a triangle for ${\cal J}_1 >0$, where the ground state is the 2-in-1-out/1-in-2-out configurations for all of the parameter space shown.  There is some mixing to the separable all-in/all-out states as a function of 
${\cal J}_2 - {\cal J}_3$, and the extremely small variation of the concurrence away from the origin is due to this mixing.  At the very centre of all the plots the ground state is exactly 4-fold degenerate, and the density is separable, resulting in zero concurrence.  The degeneracy is lifted whenever any of the parameters ${\cal J}_{2}$, ${\cal J}_3$ or ${\cal J}_4$ becomes non-zero, and a continuous increase in the concurrence from zero occurs as long as the temperature is finite.  At $T=0$, the concurrence changes discontinuously from zero to non-zero, except when varied along the line ${\cal J}_2 = {\cal J}_3$, along which the concurrence changes continuously, even at $T=0$.


\section{Discussion}
One of the motivations behind this work is to track the evolution of a highly degenerate ground state resulting from frustration into an entangled state arising from quantum fluctuations.  The most famous example of this is quantum spin ice materials, which are spin systems where the spins are arranged on the vertices of a corner-sharing tetrahedral lattice.  A `spin-ice' state occurs when coupling between the spins favours `2-in-2-out' spin configurations - configurations with   exactly two spins pointing straight into and two spins pointing straight out of each tetrahedron, a highly degenerate, classical ({\em i.e.}, separable) state. It is thought that  perturbations to the system will lead to a partial lifting of the degeneracy of the spin ice states, resulting in a highly entangled ground state \cite{rossPhysRevX.1.021002, Chen2025-PhysRevB.111.014436}, giving a change in concurrence that is discontinuous at $T=0$. In order to completely understand the transition the concurrence should be evaluated at finite temperature, in order to properly include all of the low-lying excited but nearly degenerate states. 

The triangle results discussed in Sec.\ 5.2
provide some insight about this transition at finite temperature.  For ${\cal J}_1  =1$ the ground state is the 6-fold degenerate and separable.  The very low temperature results shown on the left column of Fig.\ \ref{fi:triangle-AF} show the abrupt transition between the separable state which exists at the origin and the entangled ground state which occurs everywhere else.  The figures on the left illustrate the smooth change in concurrence as a function of the model parameters that occurs at higher temperature - the concurrence develops when the coupling constants are of the same order as the temperature.

There are obviously some important small-size effects that occur in the triangle system.  In small systems there are fewer states altogether, and in particular there are fewer states belonging to each irreducible representation.  In the triangle system, two out of three of the 
2-in-1-out/1-in-2-out doublets are the only states in their representation, while the third doublet is in a superposition with the separable all-in-all-out doublet. This superposition is evident only in the very slightly darker corners of the plots in the top row of Fig.~\ref{fi:triangle-AF}.  In larger systems the number of states increases faster than the number of irreducible representations, and so all of the energy eigenstates will be superpositions of symmetrized basis states, which may tend to reduce the concurrence and expand the range of low concurrence regions.

We conclude this section with some brief remarks about the scalability of our method.  The main challenge is that our method relies on the results of exact diagonalization to find the thermal density; however by some approximation the highest energy states may be omitted, allowing for faster diagonalization.  
In computing the concurrence of each eigenstate of a system with $N$ spins,  the measure given 
in Eq.\ \ref{N-concur} involves an evaluation of a number of terms that increases as $2^N$, but this measure
could be replaced by a simpler one, such as the $I$-concurrence \cite{rungtaPhysRevA.67.012307, Chen2025-PhysRevB.111.014436}.
The number of independent elements in the density matrix will increase with $N$, and the scaling of the increase will depend on how many additional symmetries are also introduced. 
Although the size of optimization problem will increase with the number of independent density matrix elements, it remains a straightforward algebraic procedure.




\section{Conclusion}
We present a new symmetry-based method to calculate the concurrence for a symmetric bi-partite 2D mixed state. The generalization of this method 
 to larger systems is  a new concurrence measure which we denote $C_{s}$.  $C_s$  reproduces the exact algebraic result of the concurrence for bipartite systems, and finds the correct limits of the concurrence for thermal mixed states at high and low temperatures. $C_s$ is therefore a quantitative measure of the entanglement for systems with a symmetric arrangement of interacting quantum objects, such as  qubits comprising  a qudit or quantum spins, and is suitable for  states with a 
  large ground state degeneracy or near-degeneracies.

\section*{Acknowledgments}

This work was funded by NSERC.

\bibliographystyle{unsrt}
\bibliography{citations.bib}

\begin{thebibliography}{1}

\bibitem{mintert2005}
Florian Mintert, André~R.R. Carvalho, Marek Kuś, and Andreas Buchleitner.
\newblock Measures and dynamics of entangled states.
\newblock {\em Physics Reports}, 415(4):207--259, 2005.

\bibitem{UhlmannPhysRevA.62.032307}
Armin Uhlmann.
\newblock Fidelity and concurrence of conjugated states.
\newblock {\em Phys. Rev. A}, 62:032307, Aug 2000.

\bibitem{carvelloPhysRevLett.93.230501}
Andr\'e R.~R. Carvalho, Florian Mintert, and Andreas Buchleitner.
\newblock Decoherence and multipartite entanglement.
\newblock {\em Phys. Rev. Lett.}, 93:230501, Dec 2004.

\bibitem{hughston1993}
Lane~P. Hughston, Richard Jozsa, and William~K. Wootters.
\newblock A complete classification of quantum ensembles having a given density matrix.
\newblock {\em Physics Letters A}, 183(1):14--18, 1993.

\bibitem{Hill-WootersPhysRevLett.78.5022}
Sam~A. Hill and William~K. Wootters.
\newblock Entanglement of a pair of quantum bits.
\newblock {\em Phys. Rev. Lett.}, 78:5022--5025, Jun 1997.

\bibitem{WootersPhysRevLett.80.2245}
William~K. Wootters.
\newblock Entanglement of formation of an arbitrary state of two qubits.
\newblock {\em Phys. Rev. Lett.}, 80:2245--2248, Mar 1998.

\bibitem{rossPhysRevX.1.021002}
Kate~A. Ross, Lucile Savary, Bruce~D. Gaulin, and Leon Balents.
\newblock Quantum excitations in quantum spin ice.
\newblock {\em Phys. Rev. X}, 1:021002, Oct 2011.

\bibitem{Chen2025-PhysRevB.111.014436}
C.~Wei and S.~H. Curnoe.
\newblock Concurrence and entanglement on a 16-site spin-$\frac{1}{2}$ pyrochlore cluster.
\newblock {\em Phys. Rev. B}, 111:014436, Jan 2025.

\bibitem{rungtaPhysRevA.67.012307}
Pranaw Rungta and Carlton~M. Caves.
\newblock Concurrence-based entanglement measures for isotropic states.
\newblock {\em Phys. Rev. A}, 67:012307, Jan 2003.

\end{thebibliography}

\end{document}